# Rattling enhanced superconductivity in $M$V$_2$Al$_{20}$ ($M$ = Sc, Lu, Y) intermetallic cage compounds

M. J. Winiarski[1,*], B. Wiendlocha[2], M. Sternik[3], P. Wiśniewski[4], J. R. O'Brien[5], D. Kaczorowski[4], and T. Klimczuk[1,*]

[1] Faculty of Applied Physics and Mathematics, Gdansk University of Technology, Narutowicza 11/12, 80-233 Gdansk, Poland
[2] AGH University of Science and Technology, Faculty of Physics and Applied Computer Science, Aleja Mickiewicza 30, 30-059 Krakow, Poland
[3] Institute of Nuclear Physics, Polish Academy of Sciences, Radzikowskiego 152, 31-342 Krakow, Poland
[4] Institute for Low Temperature and Structure Research, Polish Academy of Sciences, PNr 1410, 50-950 Wrocław, Poland
[5] Off Grid Research, 6501 Goodwin Street, San Diego, CA, 92111, USA

**Abstract**

Polycrystalline samples of four intermetallic compounds: $M$V$_2$Al$_{20}$ ($M$ = Sc, Y, La, and Lu) were synthesized using an arc-melting technique. The crystal structures were analyzed by means of powder x-ray diffraction and Rietveld analysis, and the physical properties were studied by means of heat capacity, electrical resistivity and magnetic susceptibility measurements down to 0.4 K. For ScV$_2$Al$_{20}$, LuV$_2$Al$_{20}$, and YV$_2$Al$_{20}$, superconductivity was observed with critical temperatures $T_c$ = 1.00 K, 0.57 K, and 0.60 K, respectively. Superconductivity for the Lu compound is reported for the first time. Theoretical calculations of the electronic and phonon structures were conducted in order to analyze the superconductivity and dynamics in ScV$_2$Al$_{20}$, YV$_2$Al$_{20}$, and LuV$_2$Al$_{20}$ and to explain the lack of a superconducting transition in LaV$_2$Al$_{20}$ down to 0.4 K. The results of the experimental and theoretical studies show that all the compounds are weakly-coupled type II BCS superconductors, and reveal the importance of the $M$-atom anharmonic "rattling" modes for the superconductivity in these materials, which seem to enhance $T_c$, especially for ScV$_2$Al$_{20}$.

*81.05.Bx, 74.25.-q, 75.20.En*

## 1 Introduction

The interesting physical behavior that is often observed in cage-type structures frequently originates from the so called "rattling" effect. In this study we associate "rattling" with a localized, anharmonic, low frequency and high amplitude vibration mode, at specific crystallographic site(s), such as at the center of an oversized atomic cage. It is believed that the "rattling" effect is responsible for superconductivity in β-pyrochlore oxides $A$Os$_2$O$_6$ ($A$ = K, Rb, Cs) [1,2], the enhancement of the thermoelectric figure of merit in skutterudite and clathrate-based semiconducting thermoelectrics [3,4] and the enhancement of the Sommerfeld coefficient (effective mass) in $SmOs_4Sb_{12}$ [5] and $PrOs_4Sb_{12}$ [6].

One of the most extensively studied systems is the β-pyrochlore $AOs_2O_6$. This family has attracted much attention due to its relatively high superconducting transition temperatures, ($T_c$ = 9.6 K for $A$ = K, $T_c$ = 6.4 K for $A$ = Rb, and $T_c$ = 3.25 K for $A$ = Cs [7]) which are clearly linked to the different ionic radii of the rattling ions: the highest $T_c$ is observed for the compound with the smallest alkali cation ($KOs_2O_6$), for which the largest vibrational anharmonicity was found in calculations [8]. On the other hand, inelastic neutron scattering measurements done by Mutka *et al.* shown that the dynamics of group I cation in $AOs_2O_6$ is more complicated than a simple single-particle "rattling" [9]. The electron-phonon coupling in $KOs_2O_6$ was found to be exceptionally strong ($\lambda_{e-p} \approx 2.4$) [10]. All these findings stimulated many theoretical investigations (see eg. refs. 11,12,13,14,15). A summary and review of experimental and theoretical studies on "rattling" and its effect on superconductivity in β-pyrochlores can be found in Ref. 16.

A series of studies of superconducting hexa- and dodecaborides showed similar behavior to that of $AOs_2O_6$. The boron network in the dodecaboride $ZrB_{12}$ is an inert background for superconductivity, as concluded from the results of the superconducting isotope effect studies on B [17] and Zr [18]. The more than one order of magnitude higher $T_c$ observed for $ZrB_{12}$ ($T_c$= 6 K) than for $LuB_{12}$ ($T_c$= 0.44 K) was explained by weaker coupling of the Einstein phonon related to the vibration of the Lu ion to the conduction electrons [19]. Studies on clathrate superconductors $Ba_8Si_{46}$ and $Ba_{24}Si_{100}$ held by Lortz *et al.* [20] had shown that the main phonon contribution to the electron-phonon coupling arises from low-energy vibrational modes of Ba ions encaged by Si atoms.

The group of isostructural intermetallic ternaries $RCr_2Al_{20}$ ($R$ = La-Nd, Gd, Ho, Er) and $RV_2Al_{20}$ ($R$ = La-Nd, Gd) was first reported in the late 1960's. In the prototypic $CeCr_2Al_{20}$ compound, space group *Fd-3m*, chromium (Wyckoff position 16*d*) and aluminum (16*c*, 48*f*, and 96*g*) atoms form cages with large icosahedral voids occupied by cerium (8*a*). Unit cell of the $CeCr_2Al_{20}$-type structure is shown in inset of Fig. 1. The atomic positions were found to be the same as in the $Mg_3Cr_2Al_{18}$ [21] and $ZrZn_{22}$ [22,23] intermetallics described earlier. In further studies, $CeCr_2Al_{20}$-type (1-2-20) aluminides were synthesized with transition metals of groups 5 and 6 as well with titanium and manganese [24,25,26,27,28,29,30,31,32]. Depending on the transition metal, different electropositive elements (rare-earth elements, actinides, and Ca) could be introduced to the 8*a* position, with compounds containing 4*d* and 5*d* metals generally forming only with the early lanthanides (La-Sm), Ca, and U [27]. Systematic studies in the 1-2-20 group has led also to the synthesis of $RT_2Zn_{20}$ ($R$ – lanthanides, Zr, Hf, Nb and U, T – transition metals of groups 7-10) [33,34,35,36,37] and $RT_2Cd_{20}$ ($R$ - rare earth, T - Ni, Pd) [38] intermetallics.

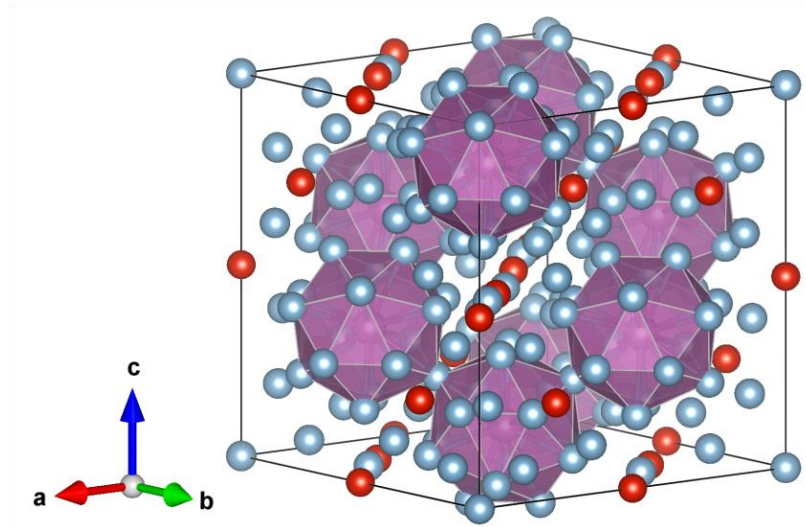

**Fig. 1** Unit cell of $CeCr_2Al_{20}$-type intermetallics. In the prototypic compound, Ce atoms (violet) occupy Wyckoff position 8*a* (⅛,⅛,⅛) surrounded by Al atoms (blue) in positions 96*g* and 16*c*. Cr atoms (red, position 16*d*) form a pyrochlore lattice [39]. More views of the $CeCr_2Al_{20}$-type structure are depicted in the Supplemental Material. Image was rendered using VESTA software [40].

Several superconductors were found among the 1-2-20 intermetallics, i.e. $Al_xV_2Al_{20}$ [41,42], $Ga_xV_2Al_{20}$, $YV_2Al_{20}$ [43], $ScV_2Al_{20}$, and $PrTi_2Al_{20}$ [44] within the aluminide group and $LaRu_2Zn_{20}$, $LaIr_2Zn_{20}$, $PrIr_2Zn_{20}$ [45], and $PrRh_2Zn_{20}$ [46] amid the Zn-based compounds. In $LaV_2Al30.1_{20}$ a strong diamagnetism was found lately but no superconducting transition was observed down to 0.4 K [47]. The strong Landau–Peierls diamagnetic susceptibility results from the peculiarities of the Fermi surface [48].

In $Al_xV_2Al_{20}$ and $Ga_xV_2Al_{20}$ the link between anharmonic rattling of Al/Ga atoms at 8*a* site and superconductivity was found [16,41,42,43]. Recently, Koza *et al.* [49,50] investigated the vibrational characteristics of several $MV_2Al_{20}$ (*M* = Sc, La, Ce, Al, Ga, and Y) by means of inelastic neutron scattering experiments and ab-initio calculations, showing the anharmonic character of *M* (8*a*) site potential in case of small cage-filling atoms (*M* = Al, Ga, Sc) and the importance of 8*a* low-energy phonon modes. On the other hand, Hasegawa *et al.* [51] and Wakiya *et al.* [52] has found by means of computational and experimental methods, respectively, that in $LaRu_2Zn_{20}$ and $LaIr_2Zn_{20}$ the Zn(16*c*) low-energy modes dominate the phonon structures.

In this study, a series of $MV_2Al_{20}$ (*M* = Sc, Y, Lu, La) cage compounds was synthesized and their physical properties were investigated by means of electrical resistivity, magnetic susceptibility and heat capacity measurements. Superconductivity was observed for $ScV_2Al_{20}$, $YV_2Al_{20}$, and $LuV_2Al_{20}$ and was not found for $LaV_2Al_{20}$ down to 0.4 K. Structural studies were performed with powder x-ray diffraction and the data were refined using the Rietveld method [53]. Finally, electronic and phonon structure calculations were carried out and the anharmonicity and electron-phonon coupling coefficients were studied in detail. The theoretical work provides a firm foundation for the interpretation of the experimental observations in terms of the differences between different rattling ions in the cages.

## 2 Experimental procedure

Samples of $M$V$_2$Al$_{20}$ were synthesized using a sequential arc-melting technique. Stoichiometric amounts of rare-earth metals (99.9% purity) and vanadium (99.8%) with approximately one third of the total aluminum (99.999%) amount were melted together in electric arc under a high purity argon atmosphere. In the second step, the remaining aluminum was added. The stepwise process was designed to ensure good homogeneity of samples with high Al content (~87 at.%). The samples were then re-melted three times to homogenize the composition. After each melting, the buttons were weighed to estimate the mass loss and small portions of Al (below 1% of the total aluminum content) were added if needed.

The as-cast samples were sealed in evacuated quartz tubes. Tantalum foil was used as a getter and to prevent a direct contact between the sample and ampoule walls. Sealed ampoules were put in a furnace for three weeks at 650°C, just below the melting point of Al (661°C). After annealing, the samples were quenched in water at room temperature.

Powder XRD measurements on powdered annealed samples were conducted on Bruker D8 diffractometer with monochromatized CuK$_{\alpha 1}$ radiation. The results were processed by means of Rietveld refinement using FULLPROF 5.30 software [54]. The initial structural models were derived from the VAl$_{10.1}$ compound structure [41].

Electrical resistivity measurements were conducted on polished samples with Pt contacts glued to the surface using a silver paste. Measurements were done in a Quantum Design Physical Property Measurement System with $^3$He cooling. Heat capacity measurements were done on small (few mg) polished samples in the same system using the standard relaxation method as well as dual slope technique [55]. Magnetic susceptibility measurements were carried out in a Quantum Design MPMS-XL SQUID magnetometer equipped with iQuantum $^3$He refrigerator.

## 3 Results

*3.1 Crystal structure*

The room temperature powder X-ray diffraction (PXRD) patterns of $M$V$_2$Al$_{20}$ ($M$ = Sc, Y, Lu, La), with a successful structural fit of the data to the CeCr$_2$Al$_{20}$ structure type, are shown in Fig. 2. The excellent quality of the refinements confirms the high purity of the materials. Only a very small amount of crystalline impurity phases were found in the ScV$_2$Al$_{20}$, YV$_2$Al$_{20}$, and LaV$_2$Al$_{20}$ samples. In case of ScV$_2$Al$_{20}$ and LaV$_2$Al$_{20}$, the impurity reflections matched the diffraction profiles of aluminum and Al$_3$V [56,57], respectively. The impurity phase in YV$_2$Al$_{20}$ could not be unambiguously identified, but most probable candidates are Al-rich binary vanadium compounds, like Al$_{23}$V$_4$ or Al$_{45}$V$_7$ [58]. It is worth noting that the Al-Y system contains a eutectic at approximately 3 at. % Y with a melting temperature of 637°C, more than 20°C below melting point of pure Al [59]. This can lead to partial melting of the sample during the annealing process at 650°C and hence result in poor material homogeneity.

The crystal structure parameters resulting from the Rietveld refinements are gathered in Table I. Refinement of the occupancy factor (SOF) of Sc, Y, La, and Lu-containing samples led to values close to 1 (a stoichiometric composition). The unit cell parameters and the dimensions of the $M$-Al polyhedra are summarized in Table II. $ScV_2Al_{20}$ has the smallest cell parameter within the series, although the relative differences between individual compounds do not exceed 1%. The unit cell dimensions of $LuV_2Al_{20}$ are lower than those of $LaV_2Al_{20}$, which is in agreement with the lanthanide contraction effect.

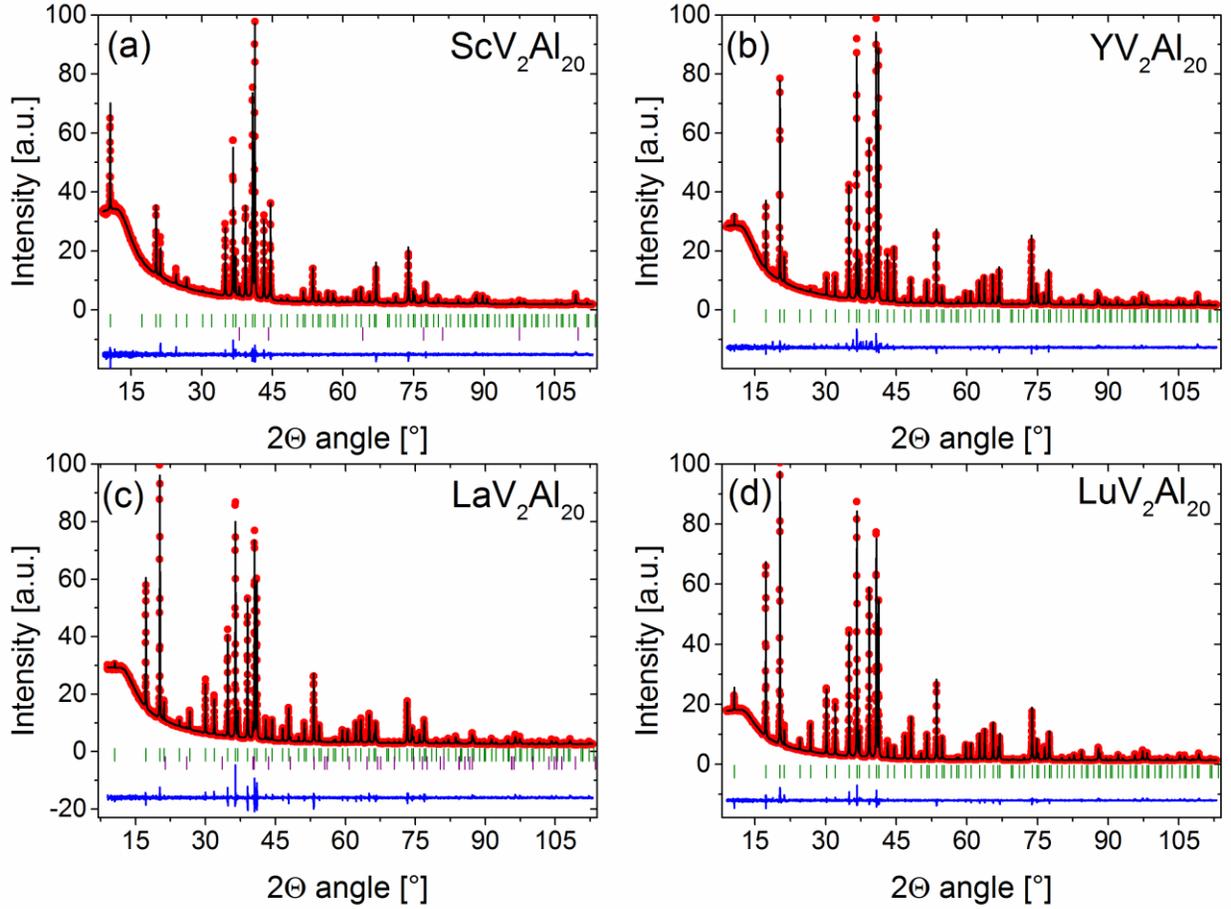

**Fig. 2** Rietveld fits to the XRD patterns obtained for (a) $ScV_2Al_{20}$, (b) $YV_2Al_{20}$, (c) $LaV_2Al_{20}$, and (d) $LuV_2Al_{20}$. Red points – observed intensities ($I_{obs}$), black line – calculated intensities ($I_{calc}$), blue line – $I_{obs}$-$I_{calc}$. Green ticks marks the expected positions of Bragg reflections for the main phase; violet – for the impurity (Al and $Al_3V$ in (a) and (c), respectively). For the values of Rietveld *R* factors and crystallographic data derived from fits see Table I.

**Table I** Crystallographic data obtained from Rietveld refinement of powder x-ray diffraction patterns. Numbers in parentheses indicate the statistical uncertainty of the least significant digit resulting from the fit. Differences in cell parameters between the compounds agree with the differences in covalent radii of the 8$a$ atoms.

|  |  | $ScV_2Al_{20}$ | $YV_2Al_{20}$ | $LaV_2Al_{20}$ | $LuV_2Al_{20}$ |
|---|---|---|---|---|---|
| Space group | | \multicolumn{4}{c}{$Fd\text{-}3m$ (# 227)} | | | |
| Pearson symbol | | \multicolumn{4}{c}{$cF184$} | | | |
| Z (number of formula units per unit cell) | | \multicolumn{4}{c}{8} | | | |
| Cell parameter (Å) | | 14.4978(1) | 14.5378(1) | 14.6219(5) | 14.5130(1) |
| Cell volume (Å$^3$) | | 3047.231 | 3072.523 | 3126.162 | 3056.832 |
| Molar weight (g·mol$^{-1}$) | | 686.4689 | 730.4189 | 780.4185 | 816.4800 |
| Density (g·cm$^{-3}$) | | 2.993 | 3.158 | 3.316 | 3.548 |
| $M$ (8$a$) | x = y = z = | \multicolumn{4}{c}{1/8} | | | |
| V (16$d$) | x = y = z = | \multicolumn{4}{c}{1/2} | | | |
| Al1 (16$c$) | x = y = z = | \multicolumn{4}{c}{0} | | | |
| Al2 (48$f$) | x = | 0.4873(2) | 0.4861(1) | 0.4864(2) | 0.4860(1) |
|  | y = z = | \multicolumn{4}{c}{1/8} | | | |
| Al3 (96$g$) | x = y = | 0.0601(1) | 0.0595(1) | 0.0588(1) | 0.0593(1) |
|  | z = | 0.3238(1) | 0.3251(1) | 0.3266(1) | 0.3245(1) |
| Figures of merit: | | | | | |
|  | $R_p$ (%) | 27.4 | 20.8 | 19.3 | 12.6 |
|  | $R_{wp}$ (%) | 16.2 | 14.1 | 13.5 | 9.61 |
|  | $R_{exp}$ (%) | 12.32 | 9.03 | 7.93 | 6.14 |
|  | $\chi^2$ (%) | 1.73 | 2.44 | 2.88 | 2.45 |

**Table II** Crystallographic data obtained from the Rietveld refinement. The *M*-Al cage radius is defined as an average bond length between the *M* atom (8*a* position) and surrounding Al atoms. Numbers given in parentheses are uncertainties of the least significant digit(s). Cage distortion indices [60], radii and volumes were calculated using VESTA software [40]. For plot of relative cage sizes and distortion indices vs. the 8*a* atom size see Supplementary Material.

|  | ScV$_2$Al$_{20}$ | YV$_2$Al$_{20}$ | LaV$_2$Al$_{20}$ | LuV$_2$Al$_{20}$ |
|---|---|---|---|---|
| Cell parameter $a$ (Å) | 14.4978(1) | 14.5378(1) | 14.6219(5) | 14.5130(1) |
| *M*-Al cage radius $R_{cage}$ (Å) | 3.1655 | 3.1908 | 3.2357 | 3.1809 |
| *M*-Al cage volume $V_{cage}$ (Å$^3$) | 91.4167 | 93.5403 | 96.7032 | 92.6832 |
| *M*-Al cage distortion index (%) | 0.421 | 0.678 | 0.973 | 0.609 |
| Covalent radius of *M* atom $r_{atom}$ (Å) [61] | 1.70(7) | 1.90(7) | 2.07(8) | 1.87(8) |
| $r_{atom}/R_{cage}$ (%) | 53.7% | 59.5% | 64.0% | 58.8% |
| $V_{atom}/V_{cage}$ (%) | 22.5% | 30.7% | 38.4% | 29.6% |

Figure 3(a) summarizes the unit cell parameters for selected *M*V$_2$Al$_{20}$ compounds and the values obtained in this study. There is a clear trend visible for the lanthanide series, namely the cubic lattice parameter systematically decreases with decreasing atom size, entirely in line with the lanthanide contraction effect. However, for smaller atoms, including Lu, Sc, Al, and Ga, the unit cell parameter does not change significantly, despite an approximately 50% increase in the covalent radius between Al and Lu. This finding suggests that below a certain atomic radius, the filling atom affects neither the size of the hosting cage nor the unit cell volume. If the dimension of the central atom is larger than the cage size, it causes cage expansion and the unit cell increases. Figure 3(b) shows the apparently linear dependencies of cage filling factors (defined as the ratio of *M* atom radius and the average *M*-Al distance as well as the ratio of *M* atom volume and the cage volume). It is worth mentioning that the Baur distortion index [60], calculated for the icosahedral *M*-Al cages, also scales with the increasing size of the filling atom.

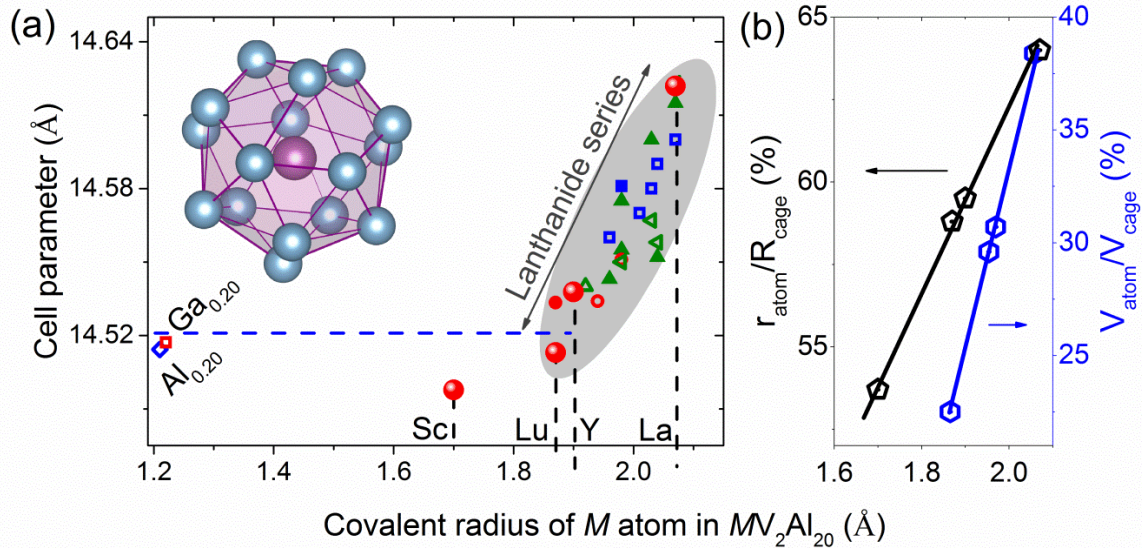

**Fig. 3 (a) Dependence of unit cell parameter, $a$, of selected $MV_2Al_{20}$ [27,41,62,63] based on the covalent radius of $M$ atom [61], $r_{atom}$. For lanthanide atoms, a linear relation between $a$ and $r_{atom}$ is observed, whereas for smaller, Al, Ga, Sc, and Lu atoms, cell parameters do not vary significantly, despite about 50% increase in the $r_{atom}$ between Al and Lu. Inset shows the $M(8a)$-Al($96g$, $16c$) cage. (b) A plot of "filling factors" (see text) dependences on the $r_{atom}$ for $MV_2Al_{20}$ ($M$ = Sc, Y, La, and Lu) compounds synthesized in this work. Both the radii (black) and volume ratios (blue) show a linear relation with $r_{atom}$ (solid lines are guide to eyes). For full version of the plot (a) with additional data and references see Supplemental Material.**

*3.2 Magnetic properties*

The temperature dependencies of the dc molar magnetic susceptibility $\chi_M(T)$ measured down to 0.5 K of $ScV_2Al_{20}$ (a), $YV_2Al_{20}$ (b) and $LuV_2Al_{20}$ (c) are shown in Figure 4. Measurements were performed after zero-field-cooling (ZFC) in an applied magnetic field of 20 Oe (40 Oe in case of $YV_2Al_{20}$). A sharp diamagnetic drop in the susceptibility (superconducting transition) is observed at 1.0 K, 0.6 K and 0.58 K for $ScV_2Al_{20}$, $YV_2Al_{20}$ and $LuV_2Al_{20}$, respectively. $\chi_M$ signal at the lowest achievable temperature ($T$ = 0.48 K) is relatively strong and almost identical for all the tested samples. For $LaV_2Al_{20}$ no superconducting transition was found down to 0.5 K, in agreement with the magnetization results reported by Onosaka, et al. [47]. To further characterize the superconducting state, the magnetization measurements versus applied field $M(H)$ were performed at T = 0.48 K. The shapes of $M(H)$ (see Fig. 4) suggest that all three compounds studied are type-II superconductors with the lower critical field $H_{c1}$(0.48 K) below 20 Oe (2 mT).

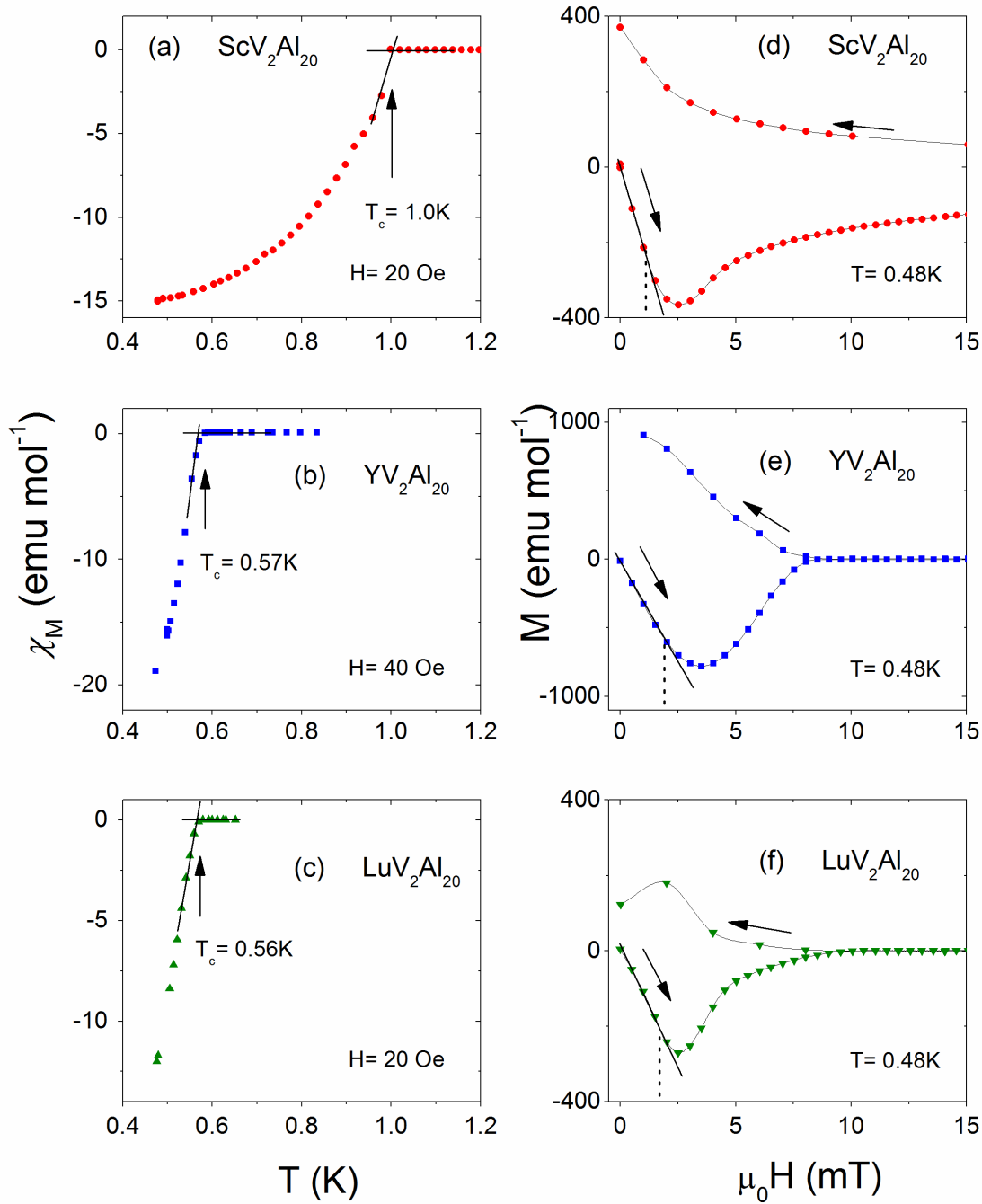

**Fig. 4 (a) – (c):** Temperature dependence of the zero-field cooled, molar magnetic susceptibility ($\chi_M$) in the vicinity of the superconducting transition for (a) $ScV_2Al_{20}$, (b) $YV_2Al_{20}$ and (c) $LuV_2Al_{20}$. (d) – (f): molar magnetization (*M*) versus applied magnetic field (*H*) measured at *T* = 0.48 K for (d) $ScV_2Al_{20}$, (e) $YV_2Al_{20}$, and (f) $LuV_2Al_{20}$.

*3.3 Electrical transport*

The temperature dependencies of the normalized resistivity ($\rho/\rho_{300K}$) of $M$V$_2$Al$_{20}$ between 0.4 K and 300 K are shown in Fig. 5. All samples exhibit metallic conductivity, and the residual resistivity ratio (*RRR* = $\rho_{300\,K}/\rho_{2\,K}$) varies from 1.1 (ScV$_2$Al$_{20}$) to 12.2 (LaV$_2$Al$_{20}$). the almost temperature independent $\rho$ observed for ScV$_2$Al$_{20}$ may arise due to either substantial scattering at the grain boundaries, or internal inhomogeneity, similar to that often observed for example in Heusler superconductors [64, 65], where heat treatment of samples can lead to enhanced antisite disorder. The second scenario is supported by the fact that the RRR increases with increasing a covalent radius of the rare earth metal *M*. Scandium is the smallest rare earth metal and due to its comparable size with vanadium, an intermixing effect in ScV$_2$Al$_{20}$ seems likely. It is worth noting that Lortz *et al.* reported that the *RRR* value for Ba$_8$Si$_{46}$ sample (with much higher $T_c$) was significantly lower than for Ba$_{24}$Si$_{100}$ [20].

Fig. 6 shows the low-temperature electrical resistivity data of ScV$_2$Al$_{20}$, YV$_2$Al$_{20}$, LaV$_2$Al$_{20}$ and LuV$_2$Al$_{20}$, measured in zero and finite external magnetic field. For ScV$_2$Al$_{20}$, YV$_2$Al$_{20}$, and LuV$_2$Al$_{20}$ a resistivity drop to zero was observed, in contrast to LaV$_2$Al$_{20}$. The superconducting critical temperatures, estimated as the midpoint of the transitions, are 1.03 K, 0.61 K and 0.60 K for the three compounds, respectively. These values are in very good agreement with the $T_c$ values estimated from the magnetic susceptibility data. The double superconducting transition for YV$_2$Al$_{20}$ that is seen for $H > 50$ Oe, suggests the presence of a second superconducting phase with a different upper critical field.

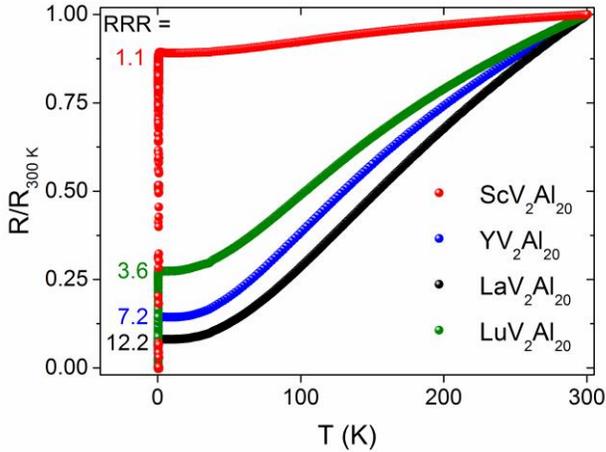

**Fig. 5 Relative resistivity ($R$(T)/$R$(300 K)) vs. temperature plots for all four materials synthesized in this study. The *RRR* ratios ($R$(300 K)/$R$(2 K)) vary from 1.1 (ScV$_2$Al$_{20}$) to 12.2 (LaV$_2$Al$_{20}$).**

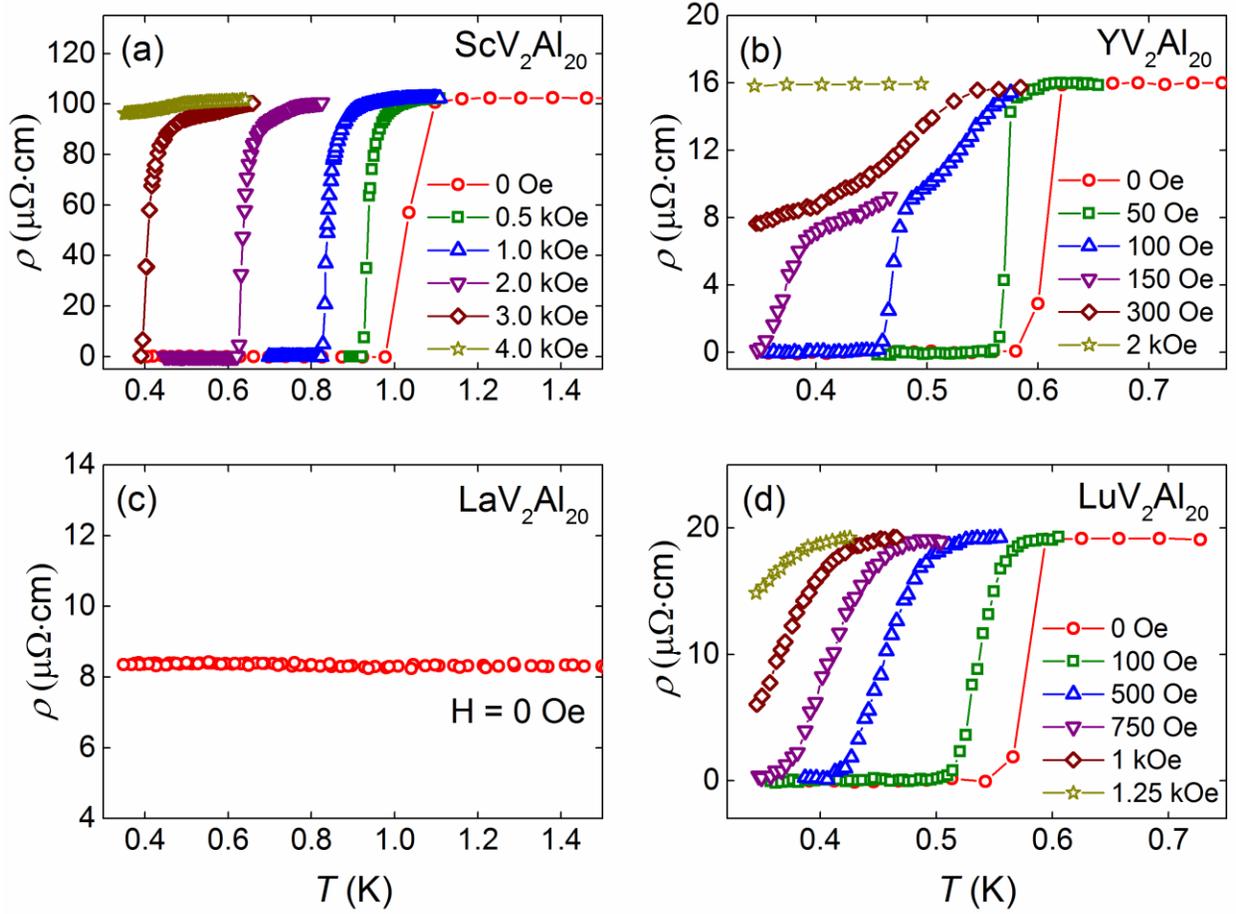

**Fig. 6** Plots of the low-temperature region of the resistivity data showing the drop to 0 at $T_c$ for (a) $ScV_2Al_{20}$, (b) $YV_2Al_{20}$, and (d) $LuV_2Al_{20}$. No superconducting transition was observed in case of (c) $LaV_2Al_{20}$. The double transition seen in plot (b) suggest the presence of material inhomogeneity.

Figure 7 presents the temperature dependence of the upper critical field for $MV_2Al_{20}$, with the data points obtained from the resistivity measurements. The initial slope $dH_{c2}/dT_c$ can be used to estimate the upper critical field $H_{c2}(0)$ using the Werthamer-Helfand-Hohenberg (WHH) expression for a dirty-limit superconductor [66]:

$$H_{c2}(0) = -0.69 \cdot T_c \left(\frac{dH_{c2}}{dT}\right)\Big|_{T=T_c} \qquad (4)$$

Taking $T_c$ obtained from the resistivity measurements, one finds $\mu_0 H_{c2}(0)$ = 0.34, 0.03 and 0.21 T for $ScV_2Al_{20}$, $YV_2Al_{20}$, and $LuV_2Al_{20}$, respectively.

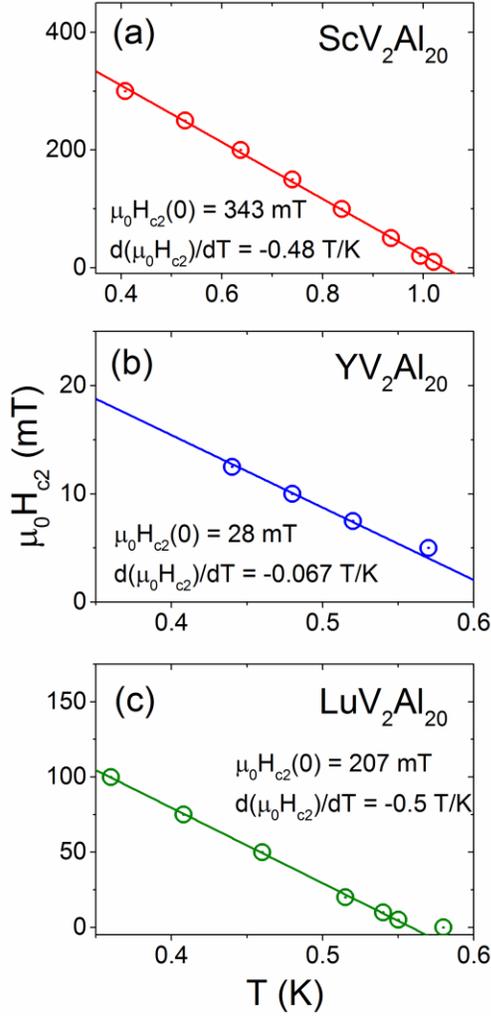

**Fig. 7 The dependence of the upper critical field on temperature for (a) ScV$_2$Al$_{20}$, (b) YV$_2$Al$_{20}$, and LuV$_2$Al$_{20}$. Solid lines are linear fits to the experimental data. The values of $\mu_0 H_{c2}(0)$ were calculated using the WHH expression (see text).**

Given the zero-temperature upper critical field $\mu_0 H_{c2}(0)$, the Ginzburg-Landau coherence length, $\xi_{GL}$, can be calculated using the formula

$$\mu_0 H_{c2} = \frac{\varphi_0}{2\pi \xi_{GL}^2} \qquad (5)$$

where $\varphi_0$ is the quantum of magnetic flux. Estimated values of $\xi_{GL}$ are 310 Å, 1084 Å, and 399 Å for Sc-, Y-, and Lu-containing samples, respectively.

*3.3 Heat capacity*

The results of heat capacity measurements for the three superconducting compounds $M$V$_2$Al$_{20}$ ($M$ = Sc, Y, Lu) are shown in Figure 8. Bulk nature of the superconductivity and good quality of the tested samples are confirmed by sharp anomalies observed at 1 K, 0.6 K and 0.57 K for ScV$_2$Al$_{20}$, YV$_2$Al$_{20}$, and LuV$_2$Al$_{20}$, respectively. The solid lines through data points represent the entropy-conservation construction, and the superconducting temperatures were estimated as a midpoint of the transition. The panels (d)-(f) of Figure 8 show the variation of $C_p/T$ with $T^2$ at zero magnetic field above the superconducting transition. The data points were fitted using the formula $\frac{C_p}{T} = \gamma + \beta T^2 + \delta T^4$ to determine the Sommerfeld coefficient $\gamma$ and phonon heat capacity coefficients $\beta$ and $\delta$. The fits give similar $\gamma$ values between 26.5(2) mJ mol$^{-1}$ K$^{-2}$ (YV$_2$Al$_{20}$) and 30.1(1) mJ mol$^{-1}$ K$^{-2}$ (ScV$_2$Al$_{20}$). Knowing $\gamma$ one can calculate the normalized heat capacity jump $\Delta C/\gamma T_c$, which characterizes the electron-phonon coupling. For ScV$_2$Al$_{20}$ and LuV$_2$Al$_{20}$, the estimated $\Delta C/\gamma T_c$ = 1.47 and 1.36 are close to the BCS value (1.43) for a weakly coupled superconductor. the significantly lower $\Delta C/\gamma T_c$ = 1.23 obtained for YV$_2$Al$_{20}$ is likely caused by the presence of impurity phases. This is in agreement with the XRD results, preliminary SEM+EDS microscopic studies and the resistivity data. The upper critical field determined for ScV$_2$Al$_{20}$ and LuV$_2$Al$_{20}$ amounts to ca. 0.3 and 0.2 T, respectively, while $\mu_0 H_{c2}(0)$ found for YV$_2$Al$_{20}$ is smaller by an order of magnitude.

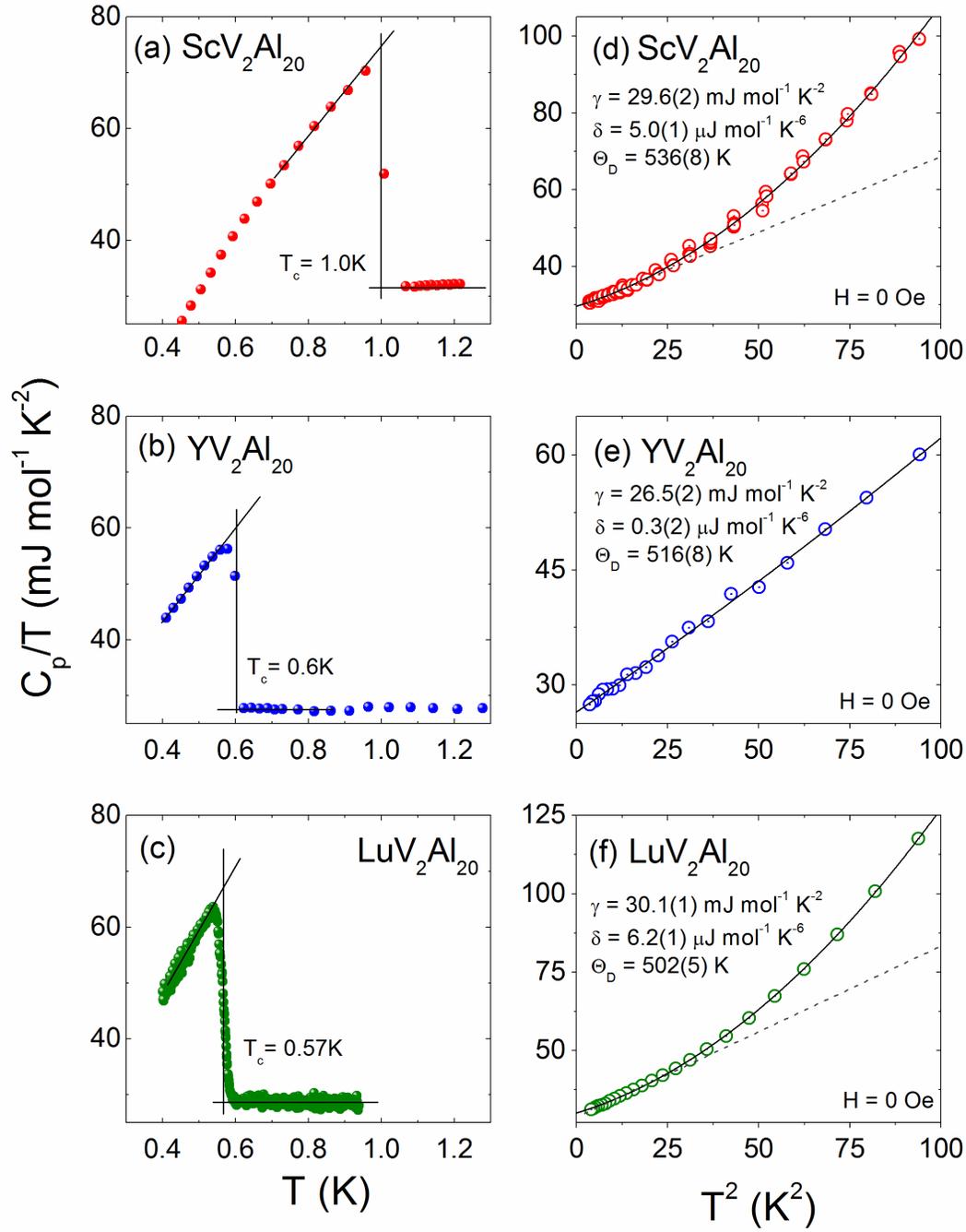

**Fig. 8** Plots of the low-temperature part of heat capacity measurements for (a) $ScV_2Al_{20}$, (b) $YV_2Al_{20}$, and (c) $LuV_2Al_{20}$, showing the anomalies due to the superconducting transition. Superconducting critical temperatures were estimated using the entropy conserving construction (black solid lines). Plots (d-f) show the fits (black solid lines) to the low temperature heat capacity results above the superconducting transition obtained at zero magnetic fields. $C_p/T$ vs. $T^2$ curves for $YV_2Al_{20}$ (e) and $LaV_2Al_{20}$ (not shown) show a linear character up to 10 K ($\sim\Theta_D/50$), whereas for the cases of $ScV_2Al_{20}$ (d) and $LuV_2Al_{20}$ (f) a clear quadratic dependence is clearly visible above ca. 5 K. Gray dashed lines (d, f) are added to emphasize the deviation from linearity. Plots shown on (d-f) are compared with the data for $LaV_2Al_{20}$ in Supplemental Material.

In a simple Debye model, the β coefficient is related to the Debye temperature $\Theta_D$ through the relation:

$$\Theta_D = \sqrt[3]{\frac{12\pi^4 nR}{5\beta}} \tag{6}$$

where $n$ is the number of atoms per formula unit and $R$ is the gas constant. Although the Debye model is rather poorly applicable for complex intermetallics (see phonon computations results in the next Section), it was applied here as a standard description of the physical properties of $MV_2Al_{20}$ compounds. The so-obtained Debye temperatures fall in the range of approximately 500 K, and thus are higher than the Debye temperature for pure Al ($\Theta_D$ = 428 K). The highest value of $\Theta_D$ = 536(8) K was found for $ScV_2Al_{20}$ that bears the lightest $M$ element, and the lowest value of $\Theta_D$ = 502(5) K was obtained for the heaviest $LuV_2Al_{20}$, as expected. This simple reasoning, however, does not apply to $YV_2Al_{20}$ and $LaV_2Al_{20}$ ($\Theta_D$ = 516(8) and 525(8) K, respectively).

The observed departures from $C_{lattice} \sim T^3$ behavior (see Fig. 8 (d,f)) signal that the density of low frequency phonon states does not follow the $\omega^2$-law that is assumed in the Debye model. In most cases, at temperatures below $\Theta_D/50$ the deviation from the $\omega^2$-law is negligible, thus $\delta T^4$ and higher terms of the Debye heat capacity expansion are significant only at higher temperatures, between approximately $\Theta_D/50$ and $\Theta_D/10$ [67]. Values of the $\delta$ heat capacity coefficient obtained for $LaV_2Al_{20}$ and $YV_2Al_{20}$ are an order of magnitude lower than those derived for $ScV_2Al_{20}$ and $LuV_2Al_{20}$. In the former two compounds the $\sim T^2$ dependency between $C_p/T$ and $T^2$ holds up to 10 K ($\sim \Theta_D/50$) in contrast with the latter ones, where deviation from a linear behavior is clearly visible above ca. 5 K. This shows that in case of $ScV_2Al_{20}$ and $LuV_2Al_{20}$ the deviations of the low frequency phonon density of states from the Debye-like shape should be significantly stronger, which was indeed clearly seen in the results of theoretical calculations (see next Section). It is worth mentioning that the $C_{lattice} \sim T^5$ dependence in low temperatures was previously found in β-pyrochlores [1,16].

Given the critical temperature $T_c$, the Debye temperature $\Theta_D$, and assuming the Coulomb pseudopotential parameter $\mu^\star$ = 0.13, the electron-phonon coupling constant can be calculated using the modified McMillan's formula [68]:

$$\lambda_{el-ph} = \frac{1.04 + \mu^* \ln\left(\frac{\Theta_D}{1.45 T_C}\right)}{(1 - 0.62\mu^*)\ln\left(\frac{\Theta_D}{1.45 T_C}\right) - 1.04} \tag{7}$$

The obtained $\lambda_{el-ph}$ constant is almost identical for the all three superconducting compounds: $\lambda_{el-ph}$ = 0.45 for $ScV_2Al_{20}$ and 0.42 for $YV_2Al_{20}$ and $LuV_2Al_{20}$. These values of $\lambda_{el-ph}$ indicate that the studied compounds are weak-coupling superconductors.

The measured and calculated thermodynamic characteristics of the $MV_2Al_{20}$ ($M$ = Sc, Y, Lu, La) compounds are summarized in Table III. The critical temperature of $YV_2Al_{20}$ obtained from the analysis of the heat capacity jump (0.60 K) is slightly lower than the previously reported value of 0.69 K, also based on $C_p$ measurements [43]. The discrepancy can be explained as a higher sample quality synthesized by Onosaka et al. [43] (likely caused by a different heat-treatment method), which is also reflected by a lower value of $\frac{\Delta C_p}{\gamma T_c}$ (1.24, compared to 1.41 in ref. 43). The obtained values of the Sommerfeld coefficient in $YV_2Al_{20}$, $LaV_2Al_{20}$, and $LuV_2Al_{20}$ are in good agreement with the previous results [47].

**Table III** Parameters of the superconducting state estimated from heat capacity, magnetization and resistivity measurements. Numbers in parentheses indicate the uncertainty of the least significant digit(s).

|  | Sc | Y | Lu | La |
|---|---|---|---|---|
| $\gamma$ (mJ mol$^{-1}$ K$^{-2}$) | 29.6(2) | 26.5(2) | 30.1(1) | 19.6(2) |
| $\beta$ (mJ mol$^{-1}$ K$^{-4}$) | 0.29(1) | 0.32(1) | 0.35(1) | 0.31(1) |
| $\Theta_D$ (K) | 536(8) | 516(8) | 502(5) | 525(8) |
| $\delta$ ($\mu$J mol$^{-1}$ K$^{-6}$) | 4.96(14) | 0.33(16) | 6.19(13) | 0.64(16) |
| $T_c$ (K) | 1.00 | 0.60 | 0.57 | - |
| $\lambda_{el\text{-}ph}$ | 0.41 | 0.39 | 0.39 | - |
| $\frac{\Delta C_p}{T_c}$ (mJ mol$^{-1}$ K$^{-2}$) | 43.3 | 32.8 | 38.8 | - |
| $\frac{\Delta C_p}{\gamma T_c}$ | 1.46 | 1.24 | 1.29 | - |
| $\frac{d\mu_0 H_{c2}}{dT_c}$ (T K$^{-1}$) | -0.482(6) | -0.067(5) | -0.501(6) | - |
| $\mu_0 H_{c2}(0)$ (T) | 0.343(4) | 0.028(2) | 0.207(2) | - |
| $\xi_{GL}$ (Å) | 310(2) | 1084(39) | 399(2) | - |

## 4 Theoretical studies

To analyze the superconductivity mechanism in $M$V$_2$Al$_{20}$, Density Functional Theory (DFT) electronic structure and phonon calculations were undertaken. As the starting point, calculations for the experimental crystal structures of the studied materials were performed using the full potential linearized augmented plane wave (FP-LAPW) method, as implemented in the WIEN2k code [69]. The Perdew-Burke-Ernzerhof Generalized Gradient Approximation [70] (PBE-GGA) was used for the exchange-correlation potential, computations were done on a 10x10x10 k-point mesh and included the spin-orbit (SO) coupling. Comparison with the semi-relativistic computations showed, that even for the case of heavy $M$ elements (La, Lu) SO interaction does not modify the density of states (DOS) near the Fermi level. The total and partial atomic densities of states are gathered in Fig. 9. The overall shapes of the DOS curves are very similar in the series of compounds, which is expected due to their isoelectronic character. For the La- and Lu-containing materials, large DOS peaks due to the 4$f$ electron shells can be easily identified (see Fig. 9(c-d)). In LaV$_2$Al$_{20}$, the empty 4$f$ La states are located 3 eV above $E_F$, whereas in LuV$_2$Al$_{20}$ the DOS structure of the filled 4$f$ shell of Lu, additionally split by the spin-orbit interaction, is located 5-6 eV below $E_F$.

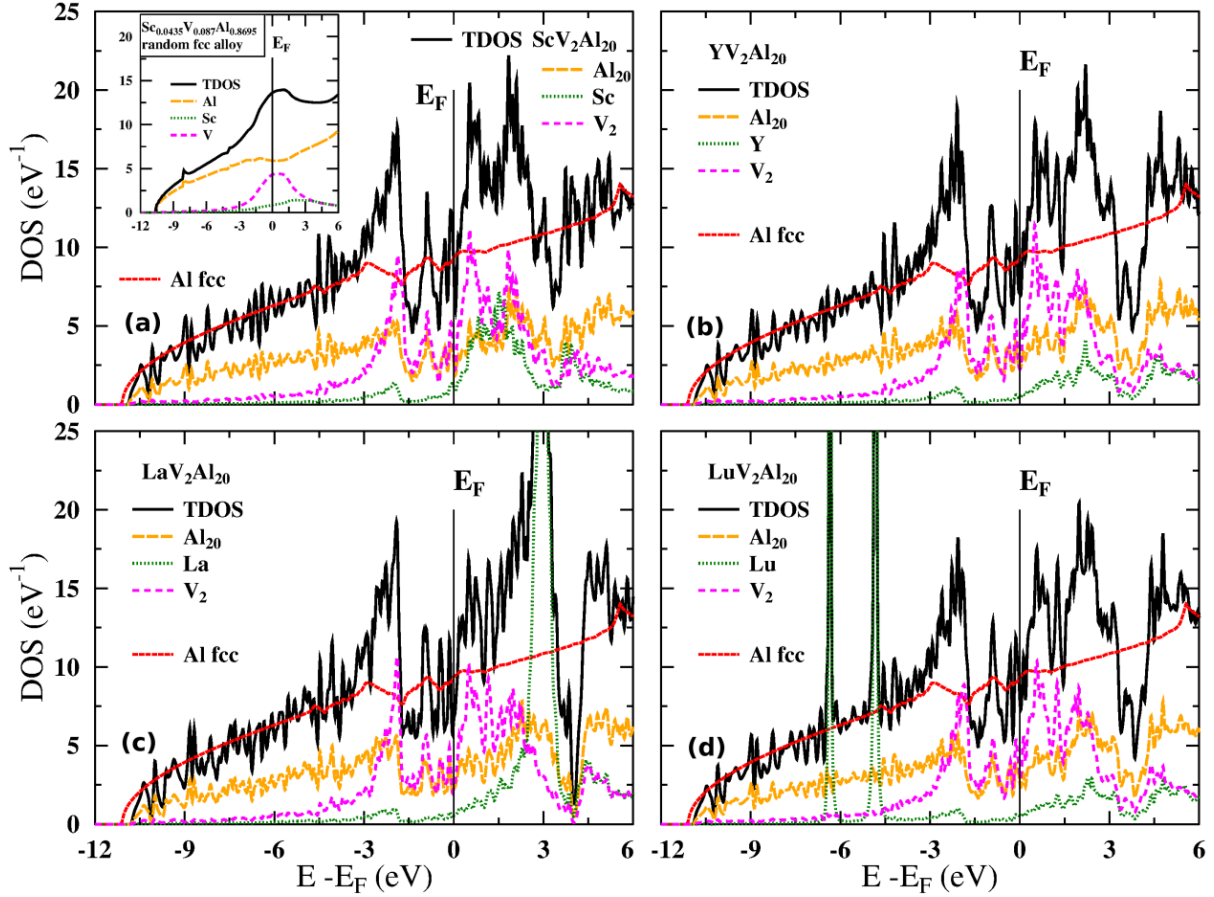

**Fig. 9** Densities of states (DOS) of the $M$V$_2$Al$_{20}$ compounds. TDOS is the total DOS per formula unit, color lines mark contributions from each group of the atoms in the unit cell (summed over all positions). The inset in (a) shows DOS of the random Sc-V-Al alloy, having equivalent composition to the ScV$_2$Al$_{20}$ compound (see, text).

It is worth noting that the envelopes of the DOS curves for the lower energies are similar to the DOS curves of metallic *fcc* aluminum, additionally plotted in the figures for convenient comparison (DOS of *fcc* Al is multiplied by 23 that is the number of atoms per f.u. in the $M$V$_2$Al$_{20}$). The significant differences start above -3 eV, where the partial DOS curves due to the M and V elements show up. This similarity might suggest that the electronic structure of $M$V$_2$Al$_{20}$ may be quite similar to the electronic structure of the (hypothetical) fcc $M$-V-Al random alloy. This supposition, however, occurred not to be true, as was verified for the $M$ = Sc example, for which additional calculations by the KKR-CPA method [71] were performed. The DOS of the random Sc-V-Al alloy is given in the inset of Fig. 9(a) and is significantly different from that of ScV$_2$Al$_{20}$. For the random alloy case, energetically unfavorable broad maximum in DOS is observed near the Fermi level, instead of a deep minimum that likely stabilizes the ordered compound.

DOS curves, expanded near the Fermi level, are presented in Fig. 10, and the corresponding values of the total DOS at $E_F$, $N(E_F)$, are shown in Table IV. In spite of the differences in the shape of DOS near $E_F$, YV$_2$Al$_{20}$, LaV$_2$Al$_{20}$ and LuV$_2$Al$_{20}$ have almost equal $N(E_F) \approx 8$ eV$^{-1}$ (per f.u.). This similarity contradicts with the experimentally observed superconducting properties, where $M$ = Y, Lu have nearly equal $T_c = 0.6$

K, whereas La-containing compound is not superconducting down to 0.4 K. In turn, $ScV_2Al_{20}$ exhibits the highest DOS value $N(E_F) = 9.0$ eV$^{-1}$, which is correlated with the highest $T_c$ within the series.

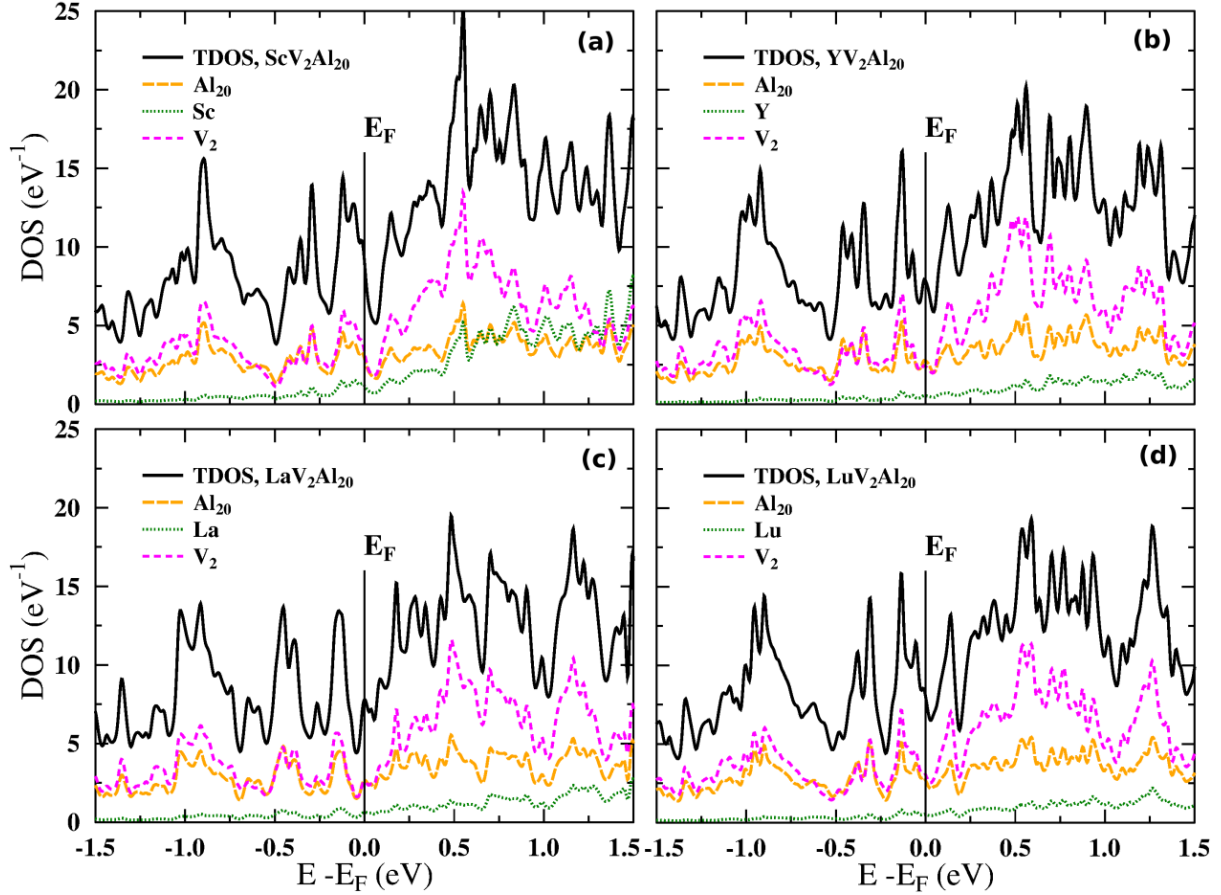

**Fig. 10 Densities of states (DOS) of the $MV_2Al_{20}$ compounds near the Fermi level. TDOS is the total DOS per formula unit, color lines mark contributions from each group of the atoms in the unit cell (summed over all positions).**

**Table IV Calculated total densities of states at the Fermi level, N(E$_F$), per formula unit, for the $MV_2Al_{20}$ compounds, corresponding Sommerfeld electronic heat capacity coefficients $\gamma_{calc}$ and electron-phonon coupling constant, derived from the comparision of the calculated and experimental ($\gamma_{expt}$) Sommerfeld parameters.**

| $M =$ | Sc | Y | La | Lu |
|---|---|---|---|---|
| $N(E_F)$ (eV$^{-1}$) per f.u. | 9.00 | 7.85 | 8.05 | 8.03 |
| $\gamma_{calc}$ (mJ mol$^{-1}$ K$^{-2}$) | 21.2 | 18.5 | 19.0 | 18.9 |
| $\lambda = \gamma_{expt}/\gamma_{calc} -1$ | 0.40 | 0.43 | 0.03 | 0.59 |

The combination of the experimentally measured electronic heat capacity coefficient $\gamma_{\text{expt}}$ with the theoretically calculated $\gamma_{\text{calc}} = \frac{\pi^2}{3} k_B^2 N(E_F)$ allows to estimate the value of the electron-phonon coupling parameter $\lambda$ from the relationship $\gamma_{\text{expt}} = \gamma_{\text{calc}}(1+\lambda)$. The computed values are presented in Table IV. For LaV$_2$Al$_{20}$, a very weak electron-phonon interaction $\lambda = 0.03$ was found, in line with non-superconducting behavior of this compound down to the lowest temperatures. The other three compounds exhibit $\lambda$ in the range $\lambda = 0.40 – 0.59$, and generally the estimation based on $\gamma$ agrees with the estimation based on $T_c$ and McMillan formula (cf. values in Tables III and IV), with the largest difference for the $M =$ Lu case ($\lambda = 0.42$ from $T_c$ versus $\lambda = 0.59$ from $N(E_F)$ and $\gamma$).

To understand the differences in the $\lambda$ and particularly the weak electron-phonon interaction in the La-bearing compound, further computations were carried out. The electronic contributions to $\lambda$, i.e. the McMillan-Hopfield parameters $\eta$ [68,72], were obtained within the Rigid Muffin Tin approximation (RMTA) [73,74,75] and KKR formalism [71] in the semi-relativistic approach. The parameters $\eta_i$ of each atom in the unit cell enter the approximate formula for the electron-phonon coupling constant:

$$\lambda = \sum_i \frac{\eta_i}{M_i \langle \omega_i^2 \rangle}, \tag{8}$$

where $\langle \omega_i^2 \rangle = \int \omega F_i(\omega) d\omega / \int \omega^{-1} F_i(\omega) d\omega$ is the "average square" phonon frequency of the atom $i$ with atomic mass $M_i$, while $F_i$ stands for the partial phonon DOS. These formulas and the RMTA approach are based on several strong assumptions (see, e.g. Refs. 74, 75), which may not be satisfied in such complicated cage compounds (see below). The computed $\eta_i$ parameters are presented in Table V, where each of the value is given per one atom, located at its crystal site in the primitive cell. Populations of each site in the primitive cell are 2 ($M$), 4 (V), 24 (Al 96$g$), 12 (Al 48$f$), 4 (Al 16$c$) (which is 1/4 of the site number, since the unit cell is of a face-centered type). First, it can be seen that the non-superconducting LaV$_2$Al$_{20}$ compound has the lowest $\eta_i$ values on all of the constituent atoms, while compared to the other three systems. The Al atoms at each of the sites exhibit similar values among the different compounds, and, due to the large population of the site, Al(96$g$) atom will be more important for the superconductivity, than Al(48$f$) and Al(16$c$). Variation in the values of $\eta_i$ at the V atom is moderate, with the smallest one seen for the LaV$_2$Al$_{20}$ case. The largest dispersion in $\eta_i$ is seen for the $M$ atoms, with the value for Sc being the highest, and that for La the lowest, more than two times smaller than that for Sc. These trends in $\eta_i$ well correspond with the measured superconducting critical temperatures. One of the reasons for the differences in the McMillan-Hopfield parameters among the isoelectronic series of $M$ elements comes from the relative amount of the partial $d$-like DOS at the Fermi level. Scandium, as the early 3$d$ element, has the largest value (~1 eV$^{-1}$ from KKR), which is next decreasing for Y and Lu (both around 0.5 eV$^{-1}$), and reaching ~0.3 eV$^{-1}$ for La, in the same order as the experimental $T_c$ value changes. The other factor, also scaling in the same way, is the lattice parameter and the cage filling factor of the $M$V$_2$Al$_{20}$, namely the smallest values are observed in ScV$_2$Al$_{20}$, similar for the compounds with Y and Lu, and the largest values for LaV$_2$Al$_{20}$. Since the McMillan-Hopfield parameters typically decrease with the increasing volume of the system (see e.g. Refs. 76, 77) this trend is reflected in the presented results, at least for the border cases of Sc- and La-bearing materials.

**Table V Computed McMillan-Hopfield parameters, $\eta_i$ in mRy/$a_B^2$, given per one atom in the primitive cell (which contains two formula units, i.e. 46 atoms). Experimental $T_c$ is repeated after Table III for convenience. All atoms are labeled with their corresponding crystal sites, the population of the site in the primitive cell is 1/4 of the site number.**

|   | $\eta_i$ (mRy/$a_B^2$) | | | | | |
|---|---|---|---|---|---|---|
| M | M(8a) | V(16d) | Al(96g) | Al(48f) | Al(16c) | $T_c$ expt. |
| Sc | 1.94 | 3.34 | 0.18 | 0.27 | 0.08 | 1.00 |
| Y | 1.53 | 3.32 | 0.18 | 0.28 | 0.08 | 0.60 |
| La | 0.86 | 2.74 | 0.15 | 0.23 | 0.06 | -- |
| Lu | 1.26 | 3.52 | 0.18 | 0.27 | 0.07 | 0.57 |

In order to perform the quantitative analysis of the electron-phonon coupling parameter and to probe the impact of atoms filling a cage on the dynamical properties of studied compounds, phonon calculations were next performed. First principles calculations were done using the projector-augmented wave formalism of the Kohn–Sham DFT within the generalized gradient approximation (GGA) approach in PAW-PBE form, implemented in the Vienna ab initio simulation package (VASP) [70,78,79]. The method of Methfessel-Paxton broadening technique with the standard 0.2 eV width of the smearing was adopted to describe the partial occupancies for each wavefunction. All calculations presented in this study were performed with a crystallographic unit cell consisting of 184 atoms. A *k*-mesh of (2,2,2) points in the Monkhorst–Pack scheme was used for the integration in reciprocal space and the energy cut-off for the plane wave expansion of 400 eV were applied. The crystal structure was optimized using the conjugate gradient technique with the energy convergence criteria set at $10^{-7}$ and $10^{-5}$ eV for the electronic and ionic iterations, respectively. The calculated lattice constants and the position of atoms are in very good agreement with the measured parameters, and are shown in Table VI. The atomic positions of *M* (1/8,1/8,1/8), V(1/2,1/2,1/2), and Al(16c)(0,0,0) were kept fixed because of the crystal symmetry.

**Table VI Relaxed crystal structures data. For experimental values see Table I.**

| $M =$ | Sc | Y | La | Lu |
|---|---|---|---|---|
| Lattice constants (Å) | 14.4391 | 14.5092 | 14.5910 | 14.4816 |
| Atomic positions: | | | | |
| Al (96g) | | | | |
| x = y | 0.0560 | 0.0590 | 0.0585 | 0.0593 |
| z | 0.3239 | 0.3251 | 0.3263 | 0.3245 |
| Al (48f) | | | | |
| x | 0.4856 | 0.4863 | 0.4871 | 0.4859 |
| y = z | ⅛ | ⅛ | ⅛ | ⅛ |
| Al (16c) | | | | |
| x = y = z | 0 | 0 | 0 | 0 |
| V (16d) | | | | |
| x = y = z | ½ | ½ | ½ | ½ |
| M (8a) | | | | |
| x = y = z | ⅛ | ⅛ | ⅛ | ⅛ |

The vibrational properties of the compounds were calculated using the direct force constant approach implemented in the program PHONON [80,81]. The force constants estimated from the first principle calculations of Hellmann–Feynman forces are used to build dynamical matrix of crystals. The phonon frequencies were obtained from the diagonalization of the dynamical matrix. In Fig. 11 the total and partial phonon density of states (PDOS) spectra calculated by the random sampling of the first Brillouin Zone at 12 000 points are presented. The total PDOS spectra cover the same frequency range up to 13 THz, and the essential differences are present at low frequencies only. For all the investigated compounds the Al(96g), Al(48f) and V(16d) partial PDOS spectra, extended on the whole frequency range, do not differ significantly. The main disparities are observed between the particular vibrational spectra of $M(8a)$ atoms and to a lesser extent by the lowest Al(16c) optic modes. The Al(16c) spectra are characterized by well defined frequencies localized below 7.5 THz, and the positions of the partial PDOS peaks fairly similar for $YV_2Al_{20}$ and $LuV_2Al_{20}$, essentially differ when compared to $ScV_2Al_{20}$ and $LaV_2Al_{20}$. The lowest Al(16c) optic mode of $LaV_2Al_{20}$ almost covers the frequency range of the La atom vibrations indicating the possible strong interaction between them. For $YV_2Al_{20}$, the overlapping of Y and Al(16c) modes is much smaller. Finally, the Sc and Lu vibrations are explicitly localized at the frequency range of acoustic modes. Thus one of the requirements imposed on rattling modes is fulfilled. Remarkably, such localized modes would have to be included to the heat capacity calculated from the Debye model, as additional Einstein modes (see below). The obtained data agree well with the experimental and theoretical results reported for $ScV_2Al_{20}$ and $LaV_2Al_{20}$ in Ref. [49], and $YV_2Al_{20}$ in Ref. [50].

The second requirement of rattling vibrations is their anharmonicity. In the figure of the dispersion relations (Fig. 12) the optical vibrations of the $M$ atoms are presented as almost dispersionless phonon modes lying in the range of the acoustic modes, around 2 THz in the case of Sc- and Lu-bearing compounds or for the lowest optic modes, around 3 THz, for Y and La ones. The $M$ atom contribution to the low-frequency modes is reflected by the partial PDOS (the middle panels of Fig. 12). In the right panels, the phonon dispersion curves calculated using three different displacements of $M$ atoms are presented in the narrow frequency range, depicted by the dashed lines in middle panels. One can see that the frequencies of a particular vibrations depend on the amplitude of the $M$ atom displacement, indicating the anharmonic character of those modes. This anharmonicity is the strongest for $M$ = Sc case, becomes weaker for Y and Lu, and is almost invisible for $M$ = La compound. At Γ point only two modes, $T_{1u}$ and $T_{2g}$, are anharmonic and both mainly consist of the $M$ atom oscillations, the highest amplitudes of remaining atoms movements are more than an order of magnitude smaller. Moreover, in $T_{2g}$ mode, no oscillations of the V and Al(3) atoms is observed. The frequency dependence of $T_{1u}$ and $T_{2g}$ on the $M$ atoms' displacement is presented in Table VII.

**Table VII** The dependence of $T_{1u}$ and $T_{2g}$ modes frequency on $M$ atoms' displacement, showing the increasing anharmonicity while going from harmonic $M$ = La, via slightly anharmonic $M$ = Y and Lu, to strongly anharmonic $M$ = Sc.

|        | $ScV_2Al_{20}$ | | $YV_2Al_{20}$ | | $LaV_2Al_{20}$ | | $LuV_2Al_{20}$ | |
|--------|-------|-------|-------|-------|-------|-------|-------|-------|
| $u$(Å) | $T_{1u}$ | $T_{2g}$ | $T_{1u}$ | $T_{2g}$ | $T_{1u}$ | $T_{2g}$ | $T_{1u}$ | $T_{2g}$ |
| 0.04   | 1.980 | 1.981 | 2.782 | 2.941 | 2.974 | 3.195 | 2.093 | 2.029 |
| 0.07   | 2.024 | 2.036 | 2.798 | 2.962 | 2.976 | 3.199 | 2.098 | 2.038 |
| 0.14   | 2.154 | 2.178 | 2.842 | 3.004 | 2.981 | 3.212 | 2.124 | 2.063 |

Analyzing the frequency of $T_{2g}$ mode, calculated for two smaller displacements, the highest changes between frequencies are found for Sc (2.7 %). For Y and Lu the differences are considerably smaller (0.7% and 0.5 %, respectively) and for La it achieves only 0.1%. It is important to point that the deviation from the harmonic oscillations related to so small displacements cannot be the quasiharmonic effects observed at high temperatures. To verify the character of this mode anharmonicity, the $T_{2g}$ mode potential energy as a function of mode amplitude $Q$ was also calculated (Fig. 13). The polarization vector of the mode allows to generate a displacement pattern corresponding to the frozen phonon. The magnitude of each atom displacement $u_i$ in a particular mode is proportional to $Q/\sqrt{m_i}$ ($m_i$ – mass of atom), e.g. the mode amplitude $Q = 14$ corresponds to the $M$ atom displacements equal 0.041 Å, 0.025 Å, 0.013 Å and 0.018 Å for Sc, Y, La and Lu. In the left panel of Fig. 13, the calculated points are presented together with two polynomial fits, purely harmonic function and harmonic with added quartic term, represented by solid and dashed lines, respectively. In the inset, the data in a whole amplitudes range used in the fitting procedure, are presented. For Sc, the potential, flat at the bottom, strongly deviates from the quadratic function. For Y and Lu the difference is not so evident. However, comparing two fits to the calculated data, the slight sign of broadening of calculated points is demonstrable, and the fits with the quartic terms are better. The $T_{2g}$ mode potential of La is purely harmonic. The flat potential, especially seen for the $M$ = Sc case, signalizes that the oscillations with large magnitude can be generated weakly changing the energy of the whole crystal. The computed mean square displacements of M atoms $<u^2>$ follows the degree of mode anharmonicity, being equal to 0.0058 (Sc), 0.0020 (Lu), 0.0015 (Y), 0.0012 (La), all in Å$^2$ and computed for 1 K, thus they follow the same order as superconducting critical temperature $T_c$.

This picture agrees very well with the analysis of atomic potential presented previously by Koza *et al.* for Sc and La [49] and Y [50]. The atomic potentials for that atoms supplemented by Lu are also presented in the right panel of Fig. 13. The atomic potentials of M atoms are shown as a dependence of total energy on the single atom displacement $u_{M(8a)}$ along (x,0,0) direction. The solid lines represent the quartic polynomial ($V(u)=au^2+bu^4$) fits to the calculated data, although the quadratic (harmonic) form to fit La atomic potential may as well be used. The largest contribution of quartic order is observed for Sc atoms what is clearly apparent in the nonlinear dependence of the restoring force (Hellmann-Feynman force) on displacement presented in the inset. Making a simple comparison of restoring forces generated on the atoms in the cage, the weak bonding of Sc atoms with the cage can be concluded. From the set of Hellmann-Feynman forces ($F(n)$) the force constants ($\Phi(n,m)$) are calculated by solving the set of equations $F(n)=\Sigma_m \Phi(n,m)u(m)$. The computed on-site force constants ($\Phi(n,n)$) for the $M$ atoms are as follows: 1.35 (Sc), 3.85 (Lu), 4.34 (Y), 7.38 (La), all in eV/Å$^2$ units, thus again, they organize themselves into three groups, correlated with the observed superconducting critical temperatures. This correlation is not surprising, since the partial electron-phonon coupling constant $\lambda_i \propto \Phi_i^{-1}$, that illustrates the important role of M atoms vibrations for the superconductivity in these materials. The smallest force constant (meaning that the smallest force is needed to move the atom) was found for the "most" superconducting ScV$_2$Al$_{20}$ case, whereas it was the largest for non-superconducting LaV$_2$Al$_{20}$. The smallest force needed to disturb atom's position also means that the response from the $M$ atom to the surrounding cage will be the smallest for Sc, so the atom is weakly coupled to its neighbors.

We can conclude that the low-frequency mode anharmonicity is clearly correlated with the magnitude of the superconductivity in $M$V$_2$Al$_{20}$ caged materials, with LaV$_2$Al$_{20}$ standing out as the only non-superconducting system, and $T_c$ enhanced in the most anharmonic ScV$_2$Al$_{20}$. The latter finding is astonishing since other caged compounds containing La, La$T_2$Zn$_{20}$ ($T$ = Ru, Ir), are superconducting. Nevertheless, when the phonon spectrum is compared to the results presented for superconducting La$T_2$Zn$_{20}$ ($T$ = Ru, Ir) by Hasegawa *et al.* [51] the major difference is seen in the 8$a$-occupying atom (La) and 16$c$ (Zn) modes. In LaRu$_2$Zn$_{20}$ the Zn(16$c$) modes clearly dominate the low-energy part of the phonon DOS separated from the relatively high-energy La(8$a$) vibrations. This apparent difference with our results can be explained by the differences in crystal structures of $M$V$_2$Al$_{20}$ and La$T_2$Zn$_{20}$. In the latter, La atoms are positioned in smaller cages and the filling factor is even higher than for LaV$_2$Al$_{20}$. Meanwhile the polyhedra centered at Zn(16$c$) are still much larger than the Zn atom, which explains the presence of low-energy modes. Also, the electronic structure of these two groups of cage compounds will be much different, thus direct comparison between them is not possible, and our conclusion of the role of anharmonicity in the superconductivity of $M$V$_2$Al$_{20}$ materials cannot be generalized to La$T_2$Zn$_{20}$.

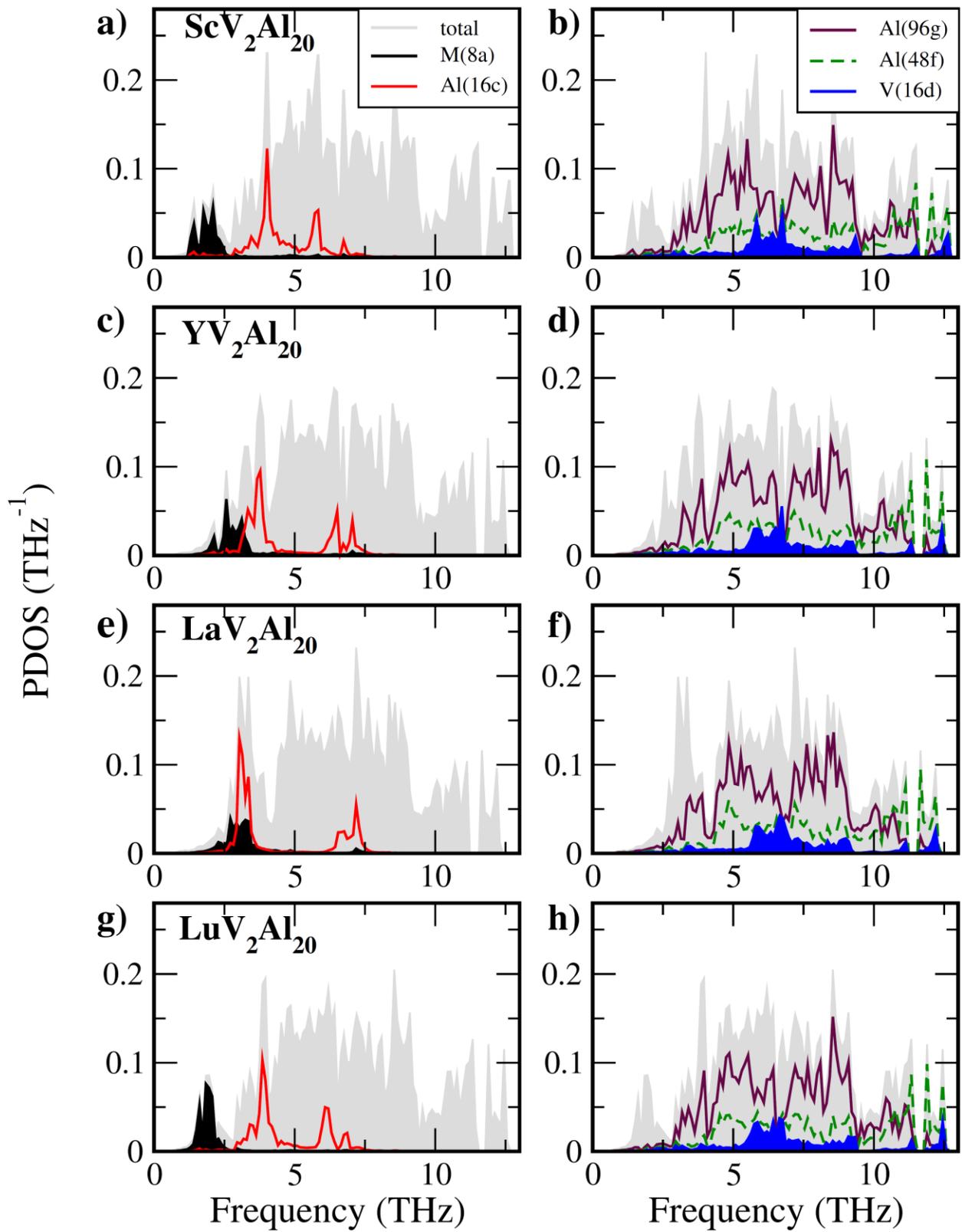

Figure 11. The total and partial phonon density of states calculated for $M$V$_2$Al$_{20}$ ($M$ = Sc, Y, La, Lu).

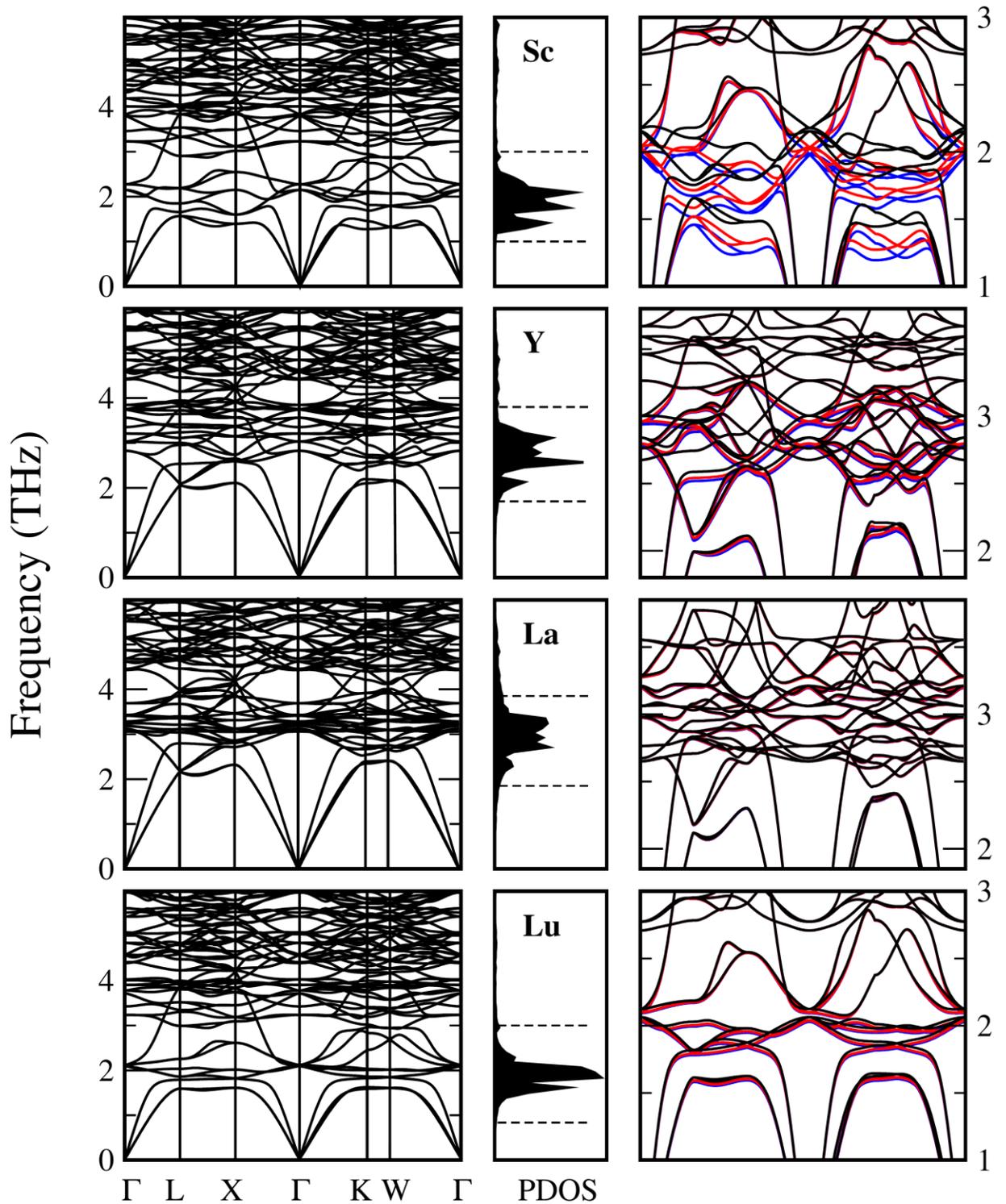

**Figure 12.** Left panels: phonon dispersion curves calculated for $M$V$_2$Al$_{20}$ ($M$ = Sc, Y, La, Lu). Middle panels: the contribution of the $M$ atoms to the phonon density of states. The dashed lines define the frequency range used in the right panels. Right panels: the dispersion curves calculated for three different displacements of $M$ atoms: 0.04 Å (blue), 0.07 Å (red) and 0.14 Å (black) presented in the frequency range of the $M$ atoms vibrations.

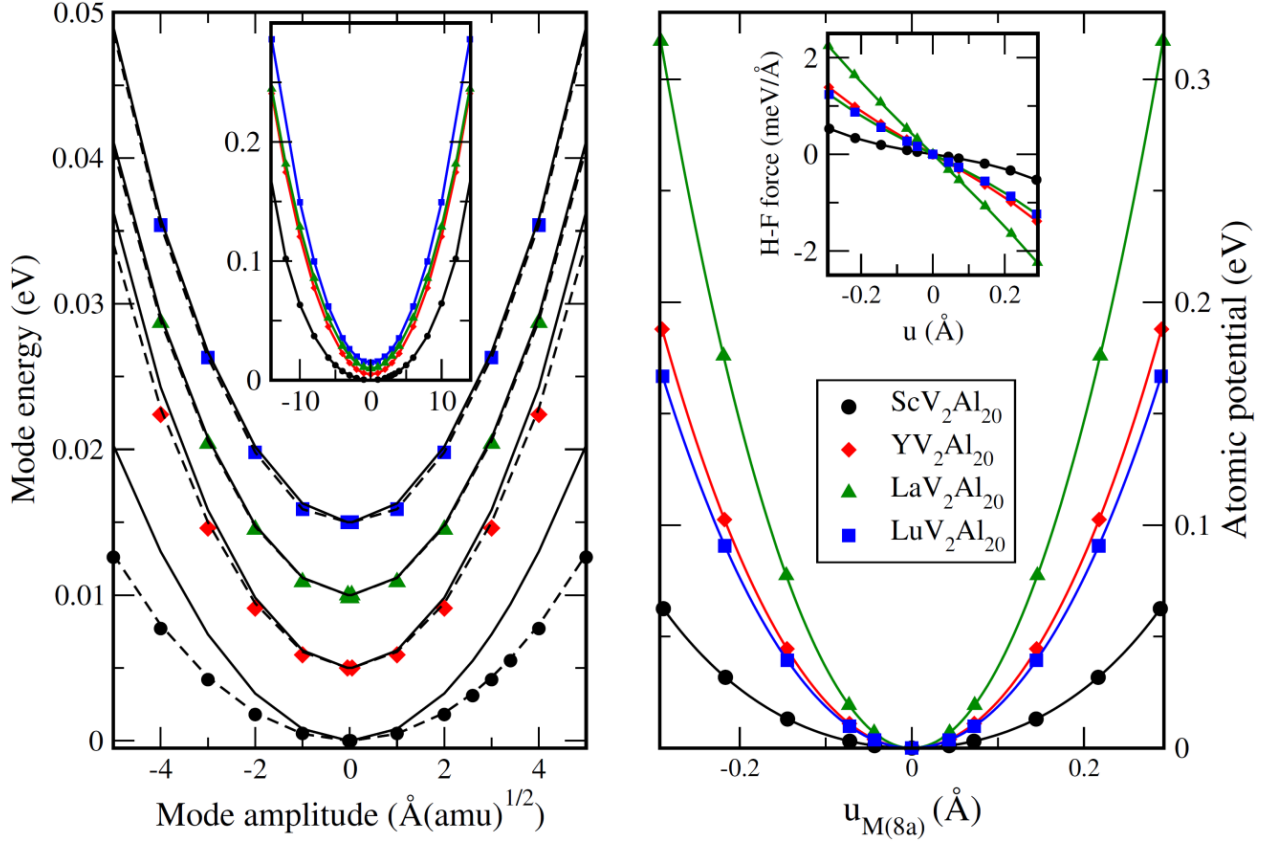

**Figure 13.** Left panel: The potential of the $T_{2g}$ mode as a function of mode amplitude in $MV_2Al_{20}$ ($M$ = Sc, Y, La, Lu). In the inset the data in a wide amplitudes range are presented. The solid and dashed lines represent, respectively, quadratic and quartic polynomial fits to the data calculated in a wide range. The whole amplitude range is shown in the inset, whereas the difference between the fits are visible in the smaller range, shown in the main panel. Right panel: The atomic potentials calculated for $M$ atoms in $MV_2Al_{20}$ ($M$ = Sc, Y, La, Lu). The inset shows the Hellmann-Feynman forces, harmonic only for the La case. Anharmonic character of the $T_{2g}$ mode potential and M atomic potentials in $M$ = Sc, Y, and Lu is visible, especially for the $M$ = Sc case.

To verify the correctness of the phonon spectra, the lattice contribution to the heat capacity was calculated, using the standard formula [82]:

$$C_{latt} = C_V = R \int_0^\infty F(\omega) \left(\frac{\hbar\omega}{k_B T}\right)^2 \frac{exp(\frac{\hbar\omega}{k_B T})}{\left(exp(\frac{\hbar\omega}{k_B T})-1\right)^2} d\omega. \qquad (9)$$

Assuming that $C_p \approx C_v$, the theoretical results can be compared to the experimental ones, as shown in Fig. 14 (the electronic contribution $\gamma T$ was added to $C_{latt}$). Agreement between the theory and the measured data is very good, even for the most critical low temperature region. This result shows that the non-linearity in the $C_p/T$ vs. $T^2$ curve for $T < 10$ K was an inherent feature of the phonon spectra of $ScV_2Al_{20}$ and $LuV_2Al_{20}$, originating from the large intensity of the low-energy optical phonon modes.

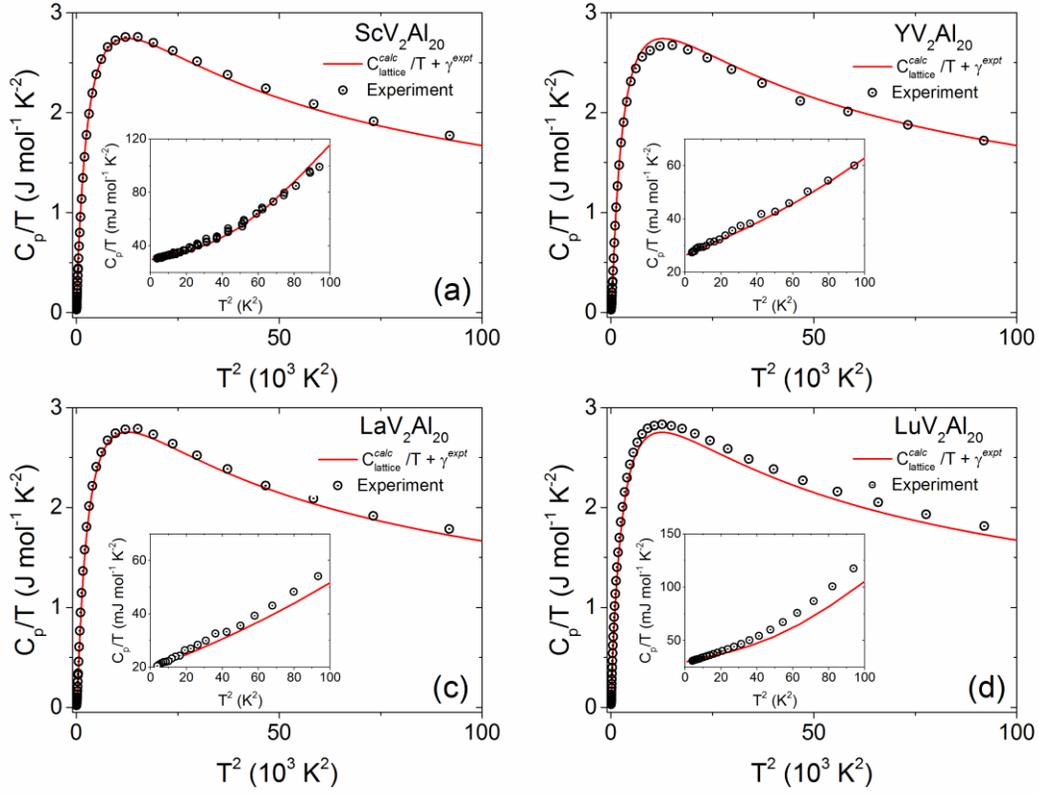

**Fig. 14. Comparison of the experimental (dots) and calculated heat capacities. The electronic contribution based on the experimental values of Sommerfeld coefficients (γ) is added to the phonon heat capacity.**

Figure 15 shows the comparison of low-temperature lattice heat capacities of the four compounds. The $\sim T^4$ upturn, correlated with the filling factor of the $M$-Al cage and observed in experimental results for $ScV_2Al_{20}$ and $LuV_2Al_{20}$, is reproduced in the calculated heat capacity.

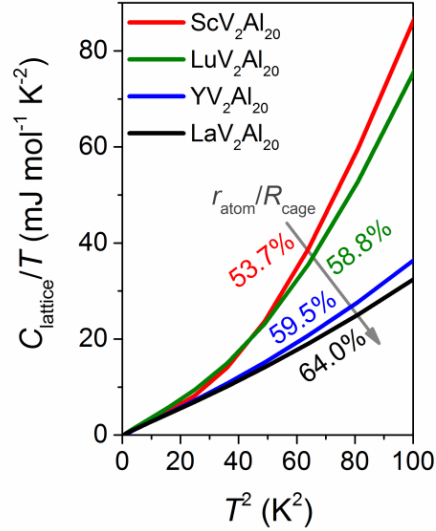

**Fig. 15. Comparison of the calculated lattice heat capacities of $MV_2Al_{20}$ ($M$ = Sc, Y, La, and Lu). Note the approximately linear character of $LaV_2Al_{20}$ and $YV_2Al_{20}$ as opposed to clearly visible upturn in case of $ScV_2Al_{20}$ and $LuV_2Al_{20}$.**

Now, using the computed McMillan-Hopfield parameters and the averaged phonon frequencies, the electron-phonon coupling constant $\lambda$ and critical temperature $T_c$ can be calculated. The superconducting critical temperature was calculated from the Allen-Dynes formula [83]:

$$T_c = \frac{\omega_{\log}}{1.20}\exp\left[\frac{-1.04(1+\lambda_{ep})}{\lambda - \mu^*(1+0.62\lambda_{ep})}\right] \qquad (10)$$

with $\omega_{\log} = \exp\left[\int F(\omega)\ln\omega\frac{d\omega}{\omega} / \int F(\omega)\frac{d\omega}{\omega}\right]$, where $F$ is the total phonon DOS (see Table VIII). Note, that the above equation was optimized for $\mu^* = 0.1$ [83] and this value was used in the theoretical computations ($\omega_{\log}$ values are considerably lower than Debye temperatures). Similarly to the dynamical properties, discussed earlier, the calculated values of $\lambda$ are forming three groups, lowest $\lambda = 0.14$ for the non-superconducting $M$ = La case, middle $\lambda = 0.20$ for $M$ = Y, Lu and highest $\lambda = 0.37$ for $M$ = Sc. These values of $\lambda$ result in the critical temperatures varying from 0 K to 0.60 K. $ScV_2Al_{20}$ has the highest calculated $T_c = 0.60$ K versus measured $T_c = 1$ K, and both $T_c$ and $\lambda$ are slightly underestimated. Qualitative agreement is observed for the non-superconducting $LaV_2Al_{20}$, which has the lowest calculated $\lambda$ and zero $T_c$. More serious underestimation is seen for $YV_2Al_{20}$ and $LuV_2Al_{20}$, the value of $\lambda = 0.20$ obtained by the RMTA approach, results in $T_c \approx 0$ K, instead of the measured $T_c \approx 0.6$ K. If the Allen-Dynes formula (10) and the computed values of $\omega_{\log}$ are used for $T_c$ calculation, $\lambda$ around 0.35 would be required to give $T_c \sim 0.5$ K, as in the experiment for those two materials. The absolute difference between the expected and calculated values of $\lambda$ is not large (0.15), but in the weak-coupling regime any small underestimations of $\lambda$ result in large deviations in $T_c$. There are several possible reasons of the underestimation of $\lambda$ connected to the main assumptions of the RMTA approach, used here (see [74,75], and references therein). First and most likely reason is the rigid ion approximation, in which it is assumed that the atomic potential moves rigidly with the vibrating atom. It works well for the transition metals and much worse for *sp*-like elements, due to delocalization of these electronic states and poor potential screening. This usually leads to the underestimation of the electron-phonon coupling parameters in such materials. As our compounds are mainly built from *p*-like Al atoms, underestimation of Al contribution of $\lambda$ in RMTA is expected, and may explain too small values of $\lambda$ among all of the superconducting $MV_2Al_{20}$ compounds. Nevertheless, the relative differences even in the underestimated values of electron-phonon coupling allow us to understand the role of each atoms' group in determining the superconducting state in this series of materials, since it is reasonable to expect, that the "background" contribution to $\lambda$ among all three superconducting materials, will be equally lowered. This is supported by almost the same values of $\eta_{Al}$ for superconducting $M$ = Sc, Y and Lu, being only lower for the non-superconducting $M$ = La. Especially, the role of $M$ atom and enhancement of $T_c$ in $ScV_2Al_{20}$, induced by the rattling-like Sc vibrations, becomes evident. Considerably lower $\lambda_{Lu/Y}$ values, when compared to $\lambda_{Sc}$, result in smaller total coupling parameter, since Al and V contributions remain similar for the three superconducting materials. This smaller $\lambda_{Lu/Y}$ is related to smaller values of the McMillan-Hopfield parameters on Lu/Y and, especially, much larger product of masses and average vibration frequencies, $(M_i\langle\omega_i^2\rangle)$ of these elements, which is in denominator in eq. (8). The effect of the decrease in mass of $M$, when going from Lu to Y and Sc, is compensated only for the Y case by the increase in the vibrational frequency ($\langle\omega_i^2\rangle$), since the interatomic force constant of Lu and Y are similar, ensuring that $M_{Lu}\langle\omega_{Lu}^2\rangle \approx M_Y\langle\omega_Y^2\rangle$. Scandium does not follow this rule, and due to much weaker coupling to the Al cage (much smaller force constant $\Phi$) the product $M_{Sc}\langle\omega_{Sc}^2\rangle$ is more than 3 times smaller, than for Lu and Y. This weaker coupling to the Al cage, which is also correlated with its anharmonic and rattling-like behavior, is then effectively increasing Sc contribution to $\lambda$, making the $T_c$ the highest among the studied series. The $M$ = La compound behaves in the opposite direction, as its perfect harmonic behavior, correlated with the strongest La-cage bonding and smallest McMillan-Hopfield parameters, remains in clear correlation with its non-superconducting behavior.

The additional reason for why the electron-phonon coupling in $ScV_2Al_{20}$ is better captured in the performed calculations, comparing to the other systems analyzed, may be related to second assumption of the RMTA approach. In the so-called *local vibration approximation*, $\lambda$ in eq. (8) is decoupled to a sum of contributions from independent crystal sites *i*, with any off-diagonal terms in $\lambda$ neglected (i.e. contributions to the electron-phonon coupling from different *i-j* sites are not considered, and in $\lambda = \sum_{ij} \lambda_{ij}$ we set $\lambda_{ij} = \lambda_i \delta_{ij}$). In our computations, the $M$ = Sc case seems to be best suited for this assumption, as well as for the "rigid-ion" one, since firstly it is an early transition metal, with the larger localization of its wave functions, while compared to 4*d* (Y) or 5*d* (Lu) elements, and secondly, Sc atoms are less coupled to the cages, due to the flat-like potential and weaker force constant. Thus, the individual contribution $\lambda_i = \eta_i / M_i \omega_i^2$ captures most of its contribution to the total coupling. When the coupling of the $M$ element to the environment becomes stronger ($\Phi$ increases), and the Y-4*d* (Lu-5*d*) wave functions become less localized, observed underestimations in $\lambda$ may suggest that the electron-phonon coupling becomes more collective property of the cage + filler and decoupling it into site contributions of the eq. (8) type (with individual atomic mass in denominator) does not work that well. Nevertheless, the performed analysis of the McMillan-Hopfield parameters and the atomic contributions to the electron-phonon coupling confirms the important role of the filler atom in determining the superconductivity of the $MV_2Al_{20}$ system, and shows the clear correlation of the low-frequency anharmonic Sc vibrations with the enhanced superconducting critical temperature in this compound, along the isoelectronic $MV_2Al_{20}$ series.

**Table VIII The average phonon energy (meV) computed using the partial phonon DOS for each of the sublattice and electron-phonon coupling parameters $\eta_i$ (atomic population of the sublattices taken into account) of $MV_2Al_{20}$ compounds, obtained using the McMillan-Hopfield parameters from Tab. V. The total $\eta$ is a sum over $i$. The value of $T_c$ is computed using the Allen-Dynes formula and $\mu^* = 0.1$.**

| $M =$ | M | V | Al(96) | Al(48) | Al(16) |
|---|---|---|---|---|---|
| Sc | \multicolumn{5}{c}{$\sqrt{\langle \omega_i^2 \rangle}$} | | | | |
|    | 8.90 | 27.57 | 26.71 | 30.71 | 18.28 |
|    | \multicolumn{5}{c}{$\lambda_i$} | | | | |
|    | 0.221 | 0.070 | 0.046 | 0.026 | 0.007 |
|    | \multicolumn{3}{l}{$\lambda$ total = 0.37, $\omega_{log}$ = 227 K} | \multicolumn{2}{l}{$T_c$ calc. = 0.60 K} | | | | |
| Y  | \multicolumn{5}{c}{$\sqrt{\langle \omega_i^2 \rangle}$} | | | | |
|    | 12.44 | 27.35 | 26.67 | 30.6 | 18.26 |
|    | \multicolumn{5}{c}{$\lambda_i$} | | | | |
|    | 0.045 | 0.071 | 0.046 | 0.027 | 0.007 |
|    | \multicolumn{3}{l}{$\lambda$ total = 0.20, $\omega_{log}$ = 245 K} | \multicolumn{2}{l}{$T_c$ calc. = $10^{-4}$ K} | | | | |
| La | \multicolumn{5}{c}{$\sqrt{\langle \omega_i^2 \rangle}$} | | | | |
|    | 13.24 | 27.63 | 26.47 | 30.48 | 17.20 |
|    | \multicolumn{5}{c}{$\lambda_i$} | | | | |
|    | 0.014 | 0.057 | 0.040 | 0.022 | 0.007 |
|    | \multicolumn{3}{l}{$\lambda$ total = 0.14, $\omega_{log}$ = 273 K} | \multicolumn{2}{l}{$T_c$ calc. = $10^{-15}$ K} | | | | |
| Lu | \multicolumn{5}{c}{$\sqrt{\langle \omega_i^2 \rangle}$} | | | | |
|    | 8.30 | 27.35 | 26.79 | 30.64 | 18.64 |
|    | \multicolumn{5}{c}{$\lambda_i$} | | | | |
|    | 0.043 | 0.075 | 0.046 | 0.026 | 0.006 |
|    | \multicolumn{3}{l}{$\lambda$ total = 0.20, $\omega_{log}$ = 228 K} | \multicolumn{2}{l}{$T_c$ calc. = $10^{-4}$ K} | | | | |

**Table IX** Values of the electron-phonon coupling parameter $\lambda_{e\text{-}p}$ obtained within different approaches: determined from the heat capacity results using relationship shown in eq. 7 (repeated from Tab. III), estimated from the comparison of experimental and theoretical values of the Sommerfeld coefficient $\gamma$ (Tab. IV), and calculated from the McMillan-Hopfield (MH) parameters and the phonon DOS (Tab. VII). For comparison, experimental and calculated values of $T_c$ are also repeated after Tabs. III and VII.

| $M =$ | Sc | Y | La | Lu |
|---|---|---|---|---|
| $\lambda_{e\text{-}p}$ | | | | |
| Calculated from heat capacity results using eq. 7 | 0.41 | 0.39 | - | 0.39 |
| Calculated from the relationship $\gamma_{expt} = \gamma_{calc}(1 + \lambda_{e-p})$ | 0.40 | 0.43 | 0.03 | 0.59 |
| Calculated from MH and the phonon DOS (see tab. VII) | 0.37 | 0.20 | 0.14 | 0.20 |
| $T_c$ (K) | | | | |
| Experimental | 1.00 | 0.60 | - | 0.57 |
| Calculated | 0.60 | $10^{-4}$ | $10^{-15}$ | $10^{-4}$ |

**Conclusions**

The results of empirical and theoretical studies of the $MV_2Al_{20}$ ($M=$ Sc, Y, La and Lu) compounds show that those with $M =$ Sc, Y, and Lu are weakly-coupled, type-II BCS superconductors with the electron-phonon coupling constant $\lambda_{el\text{-}ph.} \sim 0.4$. The estimated critical temperatures for $ScV_2Al_{20}$, $YV_2Al_{20}$, and $LuV_2Al_{20}$ (1.03, 0.61, and 0.60 K, respectively) are slightly lower than the values reported previously for $Al_xV_2Al_{20}$ (~1.5 K) and $Ga_xV_2Al_{20}$ (~1.7 K) [41,43]. Two Zn-based $CeCr_2Al_{20}$-type compounds: $LaIr_2Zn_{20}$ and $LaRu_2Zn_{20}$ show superconductivity with comparable $T_c$ (0.6 and 0.2 K, respectively) [37]. Two other examples of $CeCr_2Al_{20}$-type superconductor, $PrTi_2Al_{20}$ [44] and $PrV_2Al_{20}$ [84], exhibit rather exotic behavior, namely the coexistence of superconductivity with a quadrupolar ordering, and thus are not comparable with the compounds discussed here. Similar situation is found in $PrIr_2Zn_{20}$, where superconductivity ($T_c \approx 0.05$ K) was found to coexist with an antiferroquadrupolar order [37,85].

The electronic and phonon structure calculations confirmed the weak-coupling regime of the electron-phonon interaction in these materials, and showed the importance of the rattling-like $M$ atom modes in driving them into superconducting state. In particular, direct correlations between the critical temperature $T_c$ and the dynamical characteristics of the $M$ element (vibrational frequencies, force constants, mean square displacements) were encountered: the purely harmonic La compound was not superconducting, the weakly anharmonic Y and Lu had intermediate $T_c \sim 0.6$ K, and finally the strongly anharmonic (rattling-like) Sc compound had $T_c \sim 1$ K, considerably enhanced comparing to other materials in the studied isoelectronic series. The rigid muffin tin calculations of $\lambda$ allowed us to confirm this trend, giving the highest $\lambda$ in ScV$_2$Al$_{20}$, intermediate for $M$ = Y and Lu, and smallest for $M$ = La. The observed underestimations of $\lambda$ possibly arises because of limitations of the RMTA approach used in the theoretical computations.

The presented results underline the major influence of the cage-filling atom occupying the 8*a* position in the cubic unit cell of $M$V$_2$Al$_{20}$ on the overall electron and phonon properties of these materials. This family of intermetallic compounds enables us to cross the border between "rattling" and harmonic vibrations of the filling atom in a cage that is favorable for superconductivity, allowing us to test and model quantitatively the effect of rattling on the $T_c$ within a single structure type. This feature makes the $M$V$_2$Al$_{20}$ family interesting system to study the effect of "rattling" on superconductivity.

**Acknowledgements**

BW was partially supported by the Polish Ministry of Science and Higher Education. The research performed at the Gdansk University of Technology was financially supported by the National Science Centre (Poland) grant (DEC-2012/07/E/ST3/00584).

The authors gratefully acknowledge fruitful discussions with prof. Robert J. Cava (Princeton University)

**Supplemental Material**

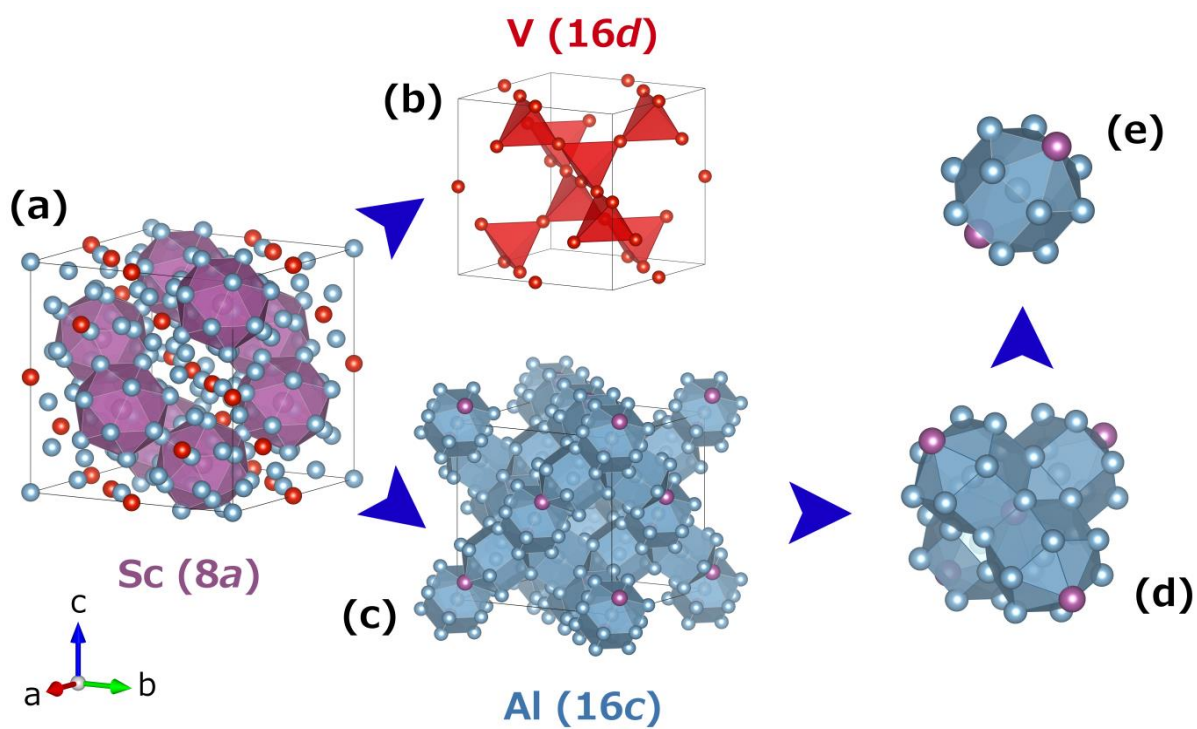

**Fig. S1** Different views of the CeCr$_2$Al$_{20}$-type structure: (a) unit cell of ScV$_2$Al$_{20}$ with diamond array of Sc-Al icosahedra (CN16 Frank-Kasper polyhedra [1]), (b) Pyrochlore lattice formed by V (16*d*) atoms. (c,d) pyrochlore lattice of Al(16*c*)-centered CN14 Frank-Kasper polyhedra. Single CN14 polyhedron composed of Al and Sc atoms is shown on (e). Image was rendered using VESTA software [2].

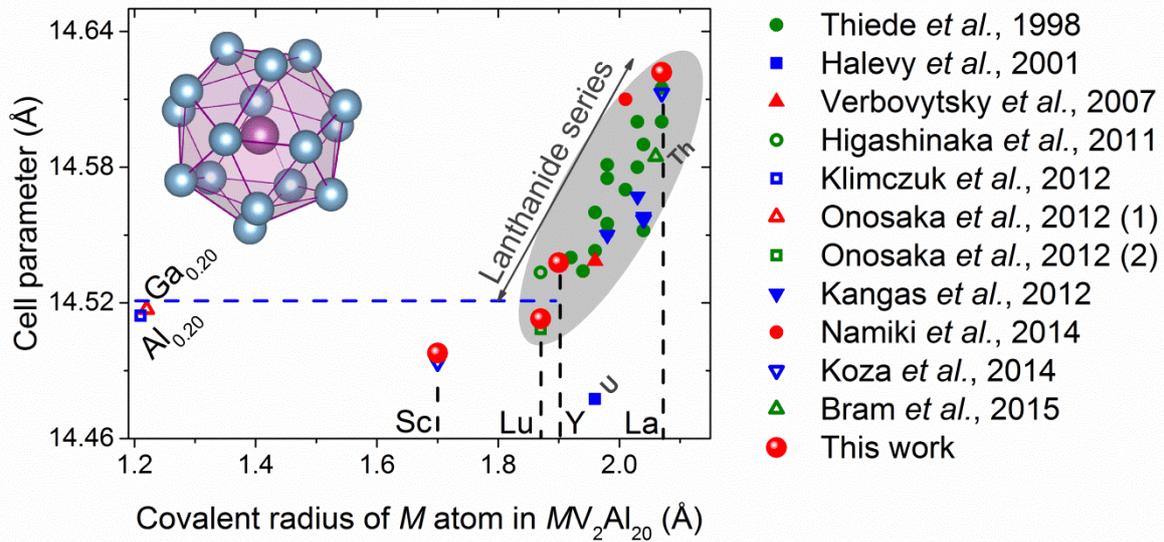

**Fig. S2** Dependence of unit cell parameter, *a*, of known $MV_2Al_{20}$ based on the covalent radius of *M* atom by Cordero *et al*. [3], $r_{atom}$. For lanthanide atoms, a linear relation between *a* and $r_{atom}$ is observed, whereas for smaller, Al, Ga, Sc, and Lu atoms, cell parameters do not vary significantly, despite about 50% increase in the $r_{atom}$ between Al and Lu. Sources of the crystallographic data: green circles – Thiede *et al*. [4] and references therein, blue squares – Halevy *et al*. [5], red triangles – Verbovytsky *et al*. [6], green squares – Higashinaka *et al*. [7], blue open squares – Klimczuk *et al*. [8], two works by Onosaka *et al*.: red triangles – 1. [9] and green open squares – 2. [10], blue triangles – Kangas *et al*. [11], red diamonds – Namiki *et al*. [12], blue open triangles – Koza *et al*.[13], and green open triangles – Bram *et al*. [14]. Two actinide-bearing compounds ($UV_2Al_{20}$ and $ThV_2Al_{20}$) were indicated by grey labels. While $ThV_2Al_{20}$ lies in the same region of the plot as lanthanide-containing materials, the lattice parameter of $UV_2Al_{20}$ clearly stands out.

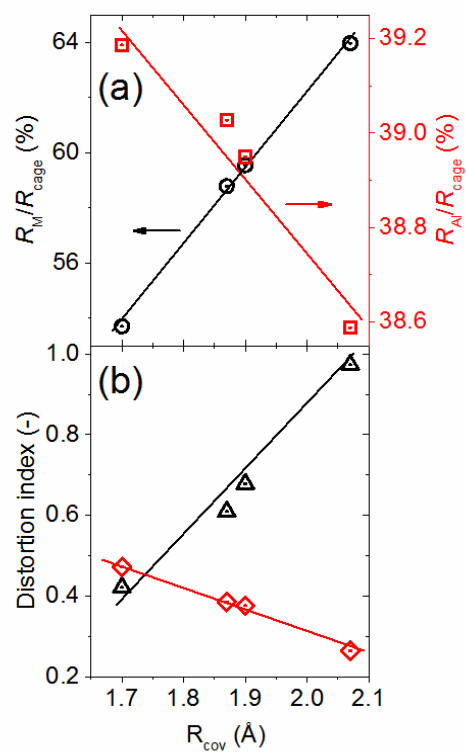

**Fig. S3** (a) The dependence of the cage filling factor, defined as the ratio of covalent radius of the filling atom to the size of the cage, on the covalent radius of $M(8a)$ atom. While the filling factor of the cage centered at $8a$ position (black circles) grow with increasing filling atom radii, the cages centered at Al($16c$) are enlarged, resulting in a decrease in their filling factor (red squares). (b) The Baur distortion index [15] for the two types of cages: $8a$-centered (black triangles) and $16c$-centered (red diamonds). The $M$-Al polyhedral are more distorted when a larger atom is put inside, as opposed to the Al($16c$) cages.

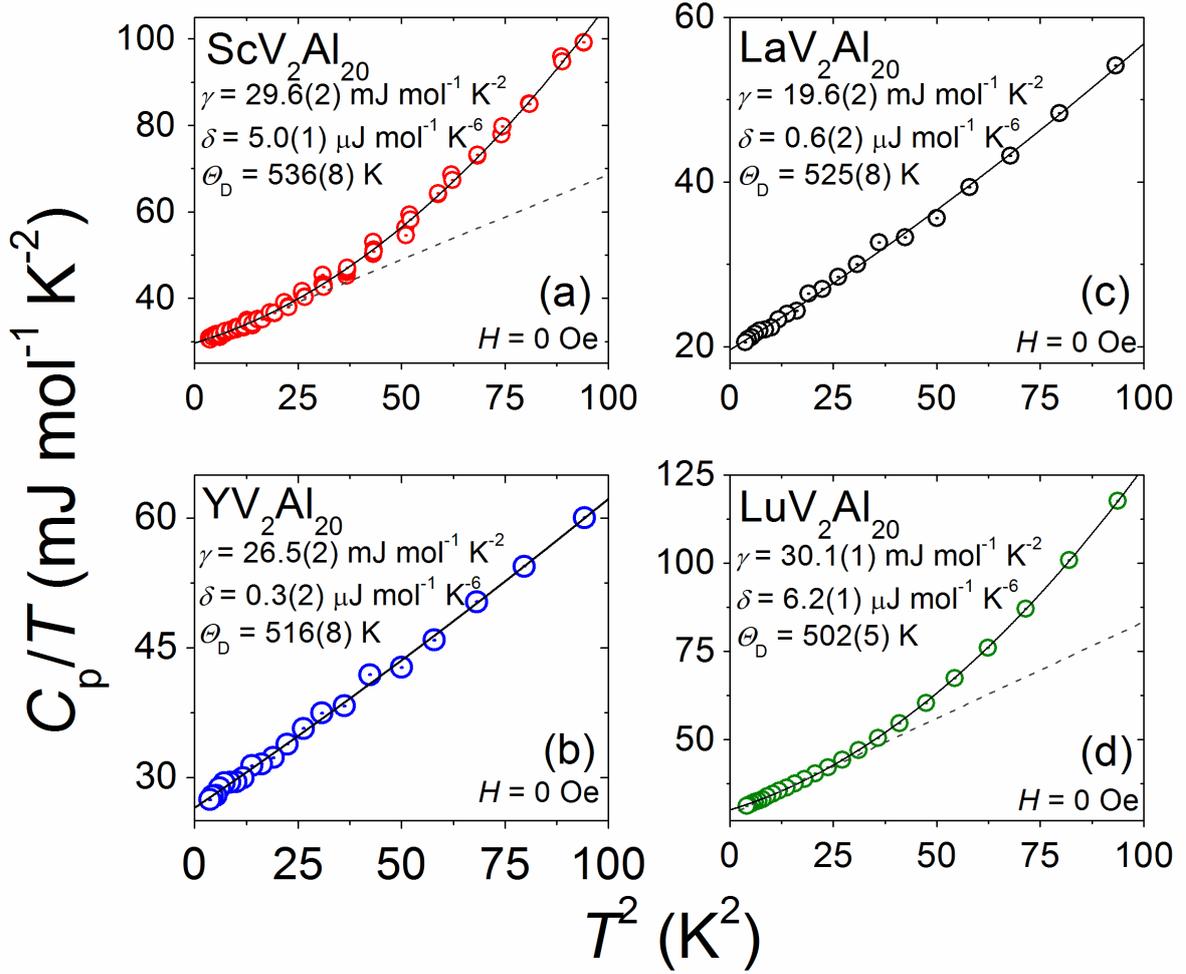

Fig. S4 Plots of $C_p/T$ vs. $T^2$ for (a) $ScV_2Al_{20}$, (b) $YV_2Al_{20}$, (c) $LaV_2Al_{20}$, and (d) $LuV_2Al_{20}$. Experimental points were fitted using formula $\frac{C_p}{T} = \gamma + \beta T^2 + \delta T^4$. In case of $ScV_2Al_{20}$ and $LuV_2Al_{20}$ (a,d) a strong deviation from linear behavior is observed in contrary to $YV_2Al_{20}$ and $LaV_2Al_{20}$ (b,c).